\documentclass[a4paper,11pt]{article}
\usepackage[utf8]{inputenc}
\usepackage{jheppub} % for details on the use of the package, please
\usepackage{bm,amsmath,amssymb,slashed,graphicx,%
            enumerate,alltt,xspace,multirow,xcolor,mathrsfs}
\usepackage{fancyvrb}
\usepackage{cancel}

\usepackage{subcaption}

\usepackage{changes}

\definechangesauthor[name=pn, color=red]{pn}

\RequirePackage{braket}
\RequirePackage{graphicx}

\RequirePackage[numbers,sort&compress]{natbib}
\makeatletter
\g@addto@macro\bfseries{\boldmath}
\makeatother

\definecolor{labelkey}{rgb}{0,0.5,0.0}

\usepackage{listings}
\definecolor{darkgreen}{rgb}{0,0.4,0}
\newcommand\Mz{M_{\scriptscriptstyle \rm Z}}

\newcommand{\mur}{\mu_{\scriptscriptstyle \rm R}}
\newcommand{\muL}{\mu_{\scriptscriptstyle \rm L}}\newcommand{\muM}{\mu_{\scriptscriptstyle \rm M}}

\newcommand\mathd{\mathrm{d}}

\newcommand{\as}{\alpha_s}

\newcommand{\muR}{\mu_{\scriptscriptstyle \rm R}}

\newcommand{\Lf}{\mathrm{Lf}}

\newcommand\rd { \mathrm{d} }

\newcommand{\Rres}{\mathcal{R}_{\rm \scriptscriptstyle log}}
\newcommand{\RresA}{\mathcal{R}_{{\rm \scriptscriptstyle log},A}}
\newcommand{\RresB}{\mathcal{R}_{{\rm \scriptscriptstyle log},B}}
\newcommand{\RresC}{\mathcal{R}_{{\rm \scriptscriptstyle log},C}}
\newcommand{\RresD}{\mathcal{R}_{{\rm \scriptscriptstyle log},D}}
\newcommand{\RresfullLL}{\mathcal{R}_{\rm \scriptscriptstyle log}^{\rm LL,full}}
\newcommand{\RresfullSPLL}{\mathcal{R}_{\rm \scriptscriptstyle log}^{\rm LL,full,SP}}
\newcommand{\RresfullNLL}{\cal{R}_{\rm \scriptscriptstyle log}^{\rm NLL,full}}

\newcommand{\beq}{\begin{eqnarray}}
\newcommand{\eeq}{\end{eqnarray}}

\newcommand{\nnb}{\nonumber}

\newcommand\Cnt{{\cal C}}

\newcommand{\Rll}{R_{\mathrm{LL}}}
\newcommand{\Rnll}{R_{\mathrm{NLL}}}
\newcommand{\Rpll}{R'_{\mathrm{LL}}}
\newcommand{\Rpnll}{R'_{\mathrm{NLL}}}

\title{On Thrust Resummation Ambiguities in $e^+e^-$ Annihilation into Hadrons}

\preprint{
\begin{flushright}
CERN-TH-2026-029
\end{flushright}
}

\author[a]{Luca Buonocore,}
\author[b]{Paolo Nason,}
\author[b,c]{Luca Rottoli,}
\author[d,e]{Paolo Torrielli\,}

\emailAdd{luca.buonocore@cern.ch}
\emailAdd{paolo.nason@mib.infn.it}
\emailAdd{luca.rottoli@unimib.it}
\emailAdd{paolo.torrielli@unito.it}

\affiliation[a]{CERN, Theoretical Physics Department, CH-1211 Geneva 23, Switzerland}
\affiliation[b]{INFN, Sezione di
  Milano-Bicocca, Piazza della Scienza 3, 20126 Milano, Italy}
\affiliation[c]{Dipartimento di Fisica G. Occhialini, Universit\`a degli Studi di Milano-Bicocca}
\affiliation[d]{Dipartimento di Fisica, Universit\`a degli Studi di Torino}
\affiliation[e]{INFN, Sezione di
  Torino, Via Giuria 1, 10125 Torino, Italy}

\date{Received: date / Accepted: \today}

\abstract{
In $e^+e^-$ shape-variable studies, and in particular for the case of thrust, fixed-order QCD predictions are typically supplemented with the resummation of contributions enhanced near the two-jet limit.

In this work we examine whether different, yet legitimate, resummation prescriptions can induce significant differences in the resulting predictions. This can occur because formally equivalent prescriptions may differ by terms that, although subleading, are characterised by asymptotic expansions and may therefore lead to slow convergence.

We first compare two alternative formulations of resummation: the conjugate-space (or Laplace-space) approach, in which resummation is performed in a variable conjugate to thrust, such that the observable factorises exactly in the soft-collinear limit; and the direct-space formulation, where resummation is instead carried out directly in the thrust variable. We show that, at double-logarithmic level, the inverse Laplace transform generates a convergent tower of subleading terms. Starting from leading-logarithmic accuracy, the expansion becomes asymptotic due to the presence of the Landau pole, leading to a mild log-factorial growth of the coefficients. When including the highest available logarithmic order in the resummation, matched to fixed-order results, we still find non-negligible differences between predictions obtained in the two spaces.

We then consider a formulation of the resummation that avoids certain approximations commonly used in the derivation of conjugate-space resummation. We observe that this also has a non-negligible numerical impact.

In general, we find that the systematics stemming from the adoption of different formalisms typically exceeds the quoted theoretical uncertainties, suggesting the need for more conservative theory-error estimates when using the thrust distribution in determinations of the strong coupling.

}


\begin{document}

\maketitle

\section{Introduction}
The electron-positron annihilation into hadrons at high energy is perhaps the
simplest framework in which to test perturbative QCD. Its advantage over
deep-inelastic scattering and hadronic collisions is due to the fact that
inclusive quantities in the asymptotic limit can be computed as functions of the
strong coupling alone. It turns out, however, that the determination of the
strong coupling constant from the distribution of shape variables in
$e^+e^- \to$ hadrons is affected by uncertainties that are, at the moment,
difficult to control.

Shape variables have been computed in fixed-order QCD calculations up to the
third power in the strong coupling constant (i.e.~N$^3$LO order)
\cite{Gehrmann-DeRidder:2007foh,Gehrmann-DeRidder:2007vsv,Gehrmann-DeRidder:2008qsl,Weinzierl:2008iv,Weinzierl:2009ms}.
Resummation of enhanced contributions near the two-jet limit is routinely
performed, using either traditional QCD resummation
methods~\cite{Catani:1992ua,Catani:1998sf,Dokshitzer:1998kz,Banfi:2001bz,Monni:2011gb,Banfi:2014sua,Tulipant:2017ybb,Banfi:2016zlc},
or Soft-Collinear Effective Theory
(SCET)~\cite{Becher:2008cf,Chien:2010kc,Becher:2012qc,Becher:2011pf}.
Furthermore it is known that non-perturbative effects cannot be neglected even
at a scale as high as LEP energies. These are treated either by using the
hadronisation models of shower Monte-Carlo generators (see for
example~\cite{Dissertori:2009ik,OPAL:2011aa,Bethke:2008hf,Dissertori:2009qa,Schieck:2012mp,Verbytskyi:2019zhh,Kardos:2018kqj}),
by using analytic models
\cite{Akhoury:1995sp,Dokshitzer:1995qm,Dokshitzer:1997iz,Dokshitzer:1998pt,Gardi:2001di,Gardi:2002bg,Gardi:2003iv,Berger:2004xf,Agarwal:2020uxi},
or by relying upon factorisation in
QCD~\cite{Korchemsky:1999kt,Korchemsky:2000kp,Bauer:2003di,Lee:2006nr}, mainly
in the SCET framework~\cite{Bauer:2000yr,Bauer:2001yt}. Determinations performed
in the analytic models typically lead to very low values of the strong coupling
constant~\cite{Abbate:2010xh,Gehrmann:2012sc,Hoang:2015hka,Benitez:2024nav,Benitez:2025vsp}.

In Refs.~\cite{Luisoni:2020efy,Nason:2023asn,Nason:2025qbx} it has been argued
that the standard approaches to the parametrisation of non-perturbative effects,
that were extrapolated from their two-jet limit value, were unjustified, and
that this extrapolation might have been the cause of the small values of $\as$
in the analytic and SCET approaches. Furthermore, it was argued there that large
uncertainties are to be associated in general with the determination of $\as$ from shape variables.

Thrust was among the first shape variables to be
introduced~\cite{Brandt:1964sa,Farhi:1977sg}, and it was also one of the first
for which the resummation of logarithmically enhanced contributions was studied.
Thrust is defined in the centre-of-mass frame of the hadronic system in terms of
the three-momenta $\vec{p}_i$ of all particles, as
\begin{equation}
T \, = \,
\max_{\vec{t}} \,
\frac{\sum_i |\vec{p}_i\cdot \vec{t}\,|}{\sum_i |\vec{p}_i|}
\, ,
\end{equation}
where $\vec{t}$ (the thrust axis) is a unit vector. One also usually defines $\tau \, \equiv \, 1-T$, and the two-jet limit is recovered when $T\to 1$, i.e.~$\tau \to 0$.

In the earliest works on thrust resummation, it was observed that the
factorisation of the phase space for soft gluon emission can be conveniently
achieved by considering the Laplace transform of the $\tau$ distribution,
thereby introducing a Laplace variable, usually denoted by $N$. Resummation was
therefore typically performed in $N$ space, and only at the end was the
distribution in $\tau$ obtained by means of an inverse Laplace transform. We
will refer to this approach as the ``conjugate'' space one. An alternative method, called ``direct'' space method, works instead directly in $\tau$ space. The two
approaches differ by terms that are subleading with respect to the declared
logarithmic accuracy of the resummation. In Ref.~\cite{Aglietti:2025jdj,Aglietti:2025ezs} it is
claimed that these subleading terms are instead numerically important, and lead to sizeable differences in the results. In particular, the fitted value of $\as$ obtained in
Ref.~\cite{Aglietti:2025jdj} is fully compatible with the PDG~average, at variance
with the results in direct space, that yield low values in tension with the
world average.\footnote{We remark that the fitting strategy
  adopted in Ref.~\cite{Aglietti:2025jdj} differs substantially from that of
  other groups. In particular, the peak region of the thrust distribution is
  included in the fit, and non-perturbative effects are modeled by means of a
  shape-function-inspired two-parameter Gaussian ansatz.}

There has been one interesting instance in the past where the choice of
resummation space made a substantial numerical difference in the value of the
predicted cross sections, i.e.~the resummation of threshold-enhanced
contributions in hadronic collisions~\cite{Catani:1996dj,Catani:1996yz}. In that
case, large differences arose between the conjugate- and the direct-space results,
because the latter contained logarithmically subleading terms with  factorially-growing coefficients. The associated perturbative expansion
was thus asymptotic, with a resummation ambiguity corresponding to an inverse
power of $Q^a$, with $a<1$, and $Q$ being the hard scale of the process. In the context of threshold resummation it was easy to show that the growth
of the perturbative coefficients in direct space is due to the violation of
longitudinal momentum conservation, which is instead conserved in conjugate
space, leading to the conclusion that the latter approach is preferable. It is
thus interesting to investigate also in the case of thrust whether there are
reasons to prefer one formulation over the other, based on the convergence
properties of the perturbative expansion.

More in general, in the present work we focus on formally subleading effects
that can in practice have an important numerical impact on the results.
In addition to those arising from the inversion from conjugate to direct space,
we consider a different mechanism that can potentially generate large subleading
terms, which we dub here the “theta-function approximation”.
This approximation has been used in the derivation of the conjugate-space
formula since the seminal paper~\cite{Catani:1992ua}, and, to the best of our
knowledge, the behaviour of the associated towers of subleading terms has
never been quantitatively examined.
Furthermore we assess whether popular numerical resummation methods~\cite{Banfi:2003je,Banfi:2001bz,Banfi:2014sua,Banfi:2018mcq,Monni:2016ktx,Bizon:2017rah}
are affected by subleading terms in similar ways.

The paper is organised as follows. In Section~\ref{sec:DL case}, we begin by
examining this aspect at the lowest logarithmic accuracy, i.e.~the
double-logarithmic (DL) level. In Section~\ref{sec:LL case}, we extend our
analysis to leading-logarithmic (LL) resummation, which differs substantially
from the double-logarithmic case because of the presence of the Landau pole. In
Section~\ref{sec:NXLL}, we examine what happens up to the highest logarithmic
accuracy currently available, namely N$^4$LL. In
Section~\ref{sec:scalevar-matching}, we study the effects of the matching
procedure used to account for known fixed-order contributions. In
Section~\ref{sec:shower-conjugate}, we analyse the impact of the theta-function
approximation employed in the derivation of the conjugate-space formula.
Finally, in Section~\ref{sec:num-resummation}, we compare the numerical
approaches CAESAR~\cite{Banfi:2003je,Banfi:2001bz},
ARES~\cite{Banfi:2014sua,Banfi:2018mcq}, and
RadISH~\cite{Monni:2016ktx,Bizon:2017rah} with the conjugate-space resummation
results. In Section~\ref{sec:conclusions}, we present our conclusions.

\section{Double-logarithmic approximation}
\label{sec:DL case}

\begin{figure}
  \begin{center}
    \includegraphics[width=0.3\textwidth]{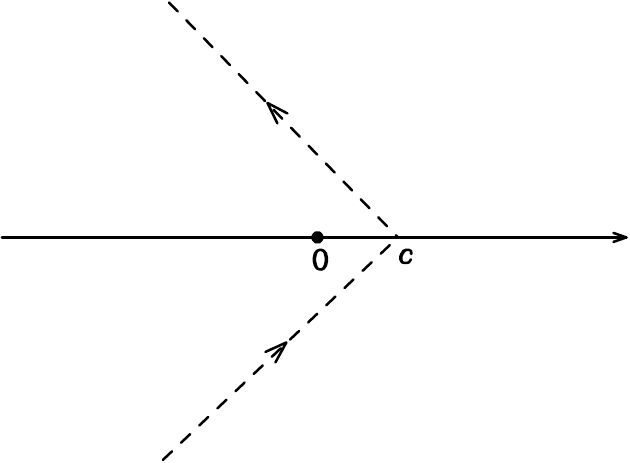}
  \end{center}
  \caption{\label{fig:Haenkel} The H\"ankel contour discussed in the text.}
\end{figure}

We begin from the resummed formula for thrust in Laplace space. At double-logarithmic (DL) level, the cumulative cross section for $1-T<\tau$ can be written as
\begin{equation}
\label{eq:cumulantMellin}
\Rres(\tau)
\, = \,
\frac{1}{2 \pi i}
\int_\Cnt
\frac{\mathd N}{N} \,
e^{N \tau} e^{- a \log^2 N}
\, ,
\end{equation}
where $a=A^{(1)} \, a_s$, $A^{(1)}=C_F$, $a_s=\alpha_s(Q)/\pi$, $Q$ is the invariant mass of the hadronic system, and $N$ is the complex Laplace variable conjugate to thrust.
The contour $\Cnt$ is a vertical straight line in the complex $N$-plane, lying at the right of all poles of the $N$ integrand. It can be parametrised as $c+ix$, with $c$ being a real positive constant, and $x$ a real parameter ranging in $(-\infty,\infty)$.
The contour $\Cnt$ can be deformed into a H\"ankel contour, parametrised as $c + ix - k|x|$, where $k$ is a positive real constant
as depicted in Fig.~\ref{fig:Haenkel}.
With this choice of contour, the convergence for large $|x|$ becomes exponential.

We assume now that Eq.~(\ref{eq:cumulantMellin}) is the exact resummation formula, i.e.~that no subleading
terms need to be added to it, and examine the convergence of its expansion in the logarithmic approximation.

The derivative of the cumulative distribution with respect to $\tau$ yields the differential distribution
\begin{equation}
\frac{\mathd \Rres(\tau)}{\mathd \tau}
\, = \,
\frac{1}{2 \pi i}
\int_\Cnt \mathd N \,
e^{N \tau} e^{- a \log^2 N}
\, ,
\end{equation}
which is well defined also for $\tau=0$, since the large-$N$ behaviour in the H\"ankel contour is suppressed by $\exp (- a \log^2 N)$.

The cumulative integral in Eq.~\eqref{eq:cumulantMellin} is dominated by values of $N \approx 1 / \tau$. It is thus convenient to introduce the variable $x = N \tau$, so that the integral in $x$ is dominated
by $x \approx 1$. We get
\beq
\label{eq:cumulantMellin2}
\Rres(\tau)
& = &
\frac{1}{2 \pi i}
\int_\Cnt \frac{\mathd N}{N} \,
e^{N \tau} e^{- a \log^2 N}
\, = \,
%\frac{1}{2 \pi i}
%\int_\Cnt \frac{\mathd x}{x} \,
%e^x e^{- a \log^2 \frac{x}{\tau}}
%\nnb\\
%& = &
\frac{1}{2 \pi i}
\int_\Cnt \frac{\mathd x}{x} \,
e^x e^{- a \left( \log x + \log \frac{1}{\tau} \right)^2}
\nnb\\
& = &
\frac{e^{- a \log^2 \frac{1}{\tau}}}{2 \pi i}
\int_\Cnt \mathd x \,
e^x e^{- \left( 1 + 2 \lambda \right)\log x  - a \log^2 x}
\, ,
% \\
%  & = & e^{- a \log^2 \frac{1}{\tau}} \exp \left( - a
%  \frac{\partial^2}{\partial \left( 1 + 2 a \log \frac{1}{\tau} \right)^2}
%  \right) \frac{1}{\Gamma \left( 1 + 2 a \log \frac{1}{\tau} \right)}
\eeq
where $\lambda = a \log \frac{1}{\tau}$.

\begin{figure}
  \begin{center}
    \includegraphics[width=0.6\textwidth]{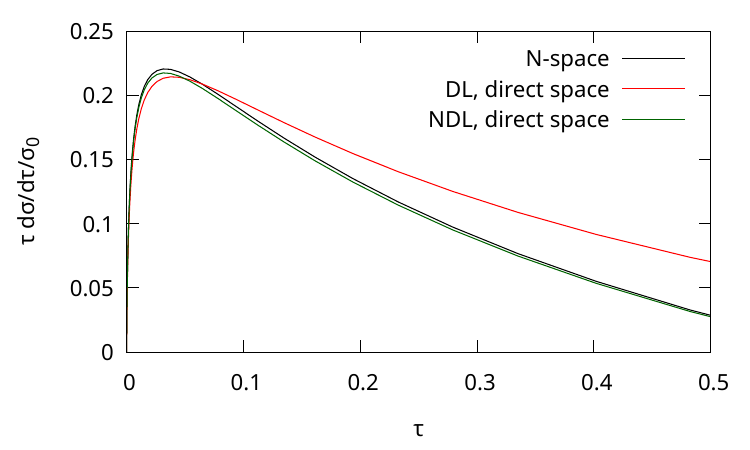}
  \end{center}
  \caption{\label{fig:DLconv} $N$-space double-log result (black), together with direct-space DL and NDL results (red and  green, respectively), obtained with $a=C_F/\pi\times 0.118$.}
\end{figure}

The perturbative expansion in $a$ at fixed
$\lambda$ in the integral in Eq.~\eqref{eq:cumulantMellin2} gives rise to the logarithmic expansion.
On the other hand,  the integral in Eq.~\eqref{eq:cumulantMellin2} is easily seen to be an analytic function of the coupling $a$ at fixed $\lambda$, for any value of $a$.
Thus, in the double-logarithmic approximation, the direct space result can be expressed
in terms of a convergent logarithmic expansion, with infinite radius of convergence in $a$.
We can write
\beq
\Rres(\tau)
& = &
\frac{e^{- a \log^2 \frac{1}{\tau}}}{2 \pi i}
\int_\Cnt \mathd x \,
e^x e^{- \log x (1 + 2 \lambda) - a \log^2 x}
\nnb\\ \label{eq:DLconjugate}
& = &
e^{- a \log^2 \frac{1}{\tau}} 
%\left[
\exp \left( - a \frac{\mathd^2}{\mathd (1 + 2 \lambda)^2} \right)
\frac{1}{\Gamma (1 + 2 \lambda)}
%\right]
\, ,
\eeq
where we have used the formula
\begin{equation}
\frac{1}{2 \pi i}
\int_\Cnt \mathd t \,
e^t e^{- z \log t} 
\, = \,
\frac{1}{\Gamma (z)}
\, .
\end{equation}
The same conclusion holds at the differential level, where we have
\beq
\frac{\mathd \Rres(\tau)}{\mathd \tau}
& = &
\frac{1}{\tau} e^{- a \log^2 \frac{1}{\tau}} 
%\left[
\exp \left( - a \frac{\mathd^2}{\mathd (2 \lambda)^2} \right)
\frac{1}{\Gamma (2 \lambda)}
%\right]
\, .
\eeq
We can define the direct-space resummed result as
\begin{equation}
\Rres^{({\rm d},n)}(\tau)
\, = \,
e^{- a \log^2 \frac{1}{\tau}} \,
\sum_{i=0}^n\frac{1}{i!} \,
\left(-a\frac{\mathd^2}{\mathd (1+2 \lambda)^2}\right)^i \,
\frac{1}{\Gamma(1+2\lambda)}
\, ,
\end{equation}
where the ``d'' in the superscript stands for ``direct''.
For $n=0$ we have the double-log (DL) approximation, for $n=1$ we have the next-to-double-log (NDL) result, and so on.
We now see that subleading terms in the relation between the two expressions are well behaved, leading to an expansion in the strong coupling that is convergent for any value of $a$.
In Fig.~\ref{fig:DLconv}
we show the $N$-space result, together with the the direct-space ones in the DL and in the NDL approximation. The convergence is clearly good.

To briefly summarise the conclusion of this section, we can simply say that a truncated expression for the
conjugate-space double-logarithmic resummation exponent leads to a direct-space exponent that has a well behaved logarithmic
expansion. We remark, however, that the inverse statement is not true, i.e.~a truncated expression
for the double-logarithmic resummation exponent in direct space leads to an  exponent in conjugate space whose logarithmic
expansion contains factorially growing terms.
In fact, if we start with the expression
\begin{equation}
{\cal R} (\tau) \, = \, \exp [- a \ln^2 \tau] \,,
\end{equation}
its Laplace transform reads
\begin{equation}
{\cal R}_N \, = \, \int_0^{\infty} \mathd \tau \, \exp [- a \ln^2 \tau] \, e^{- N \tau} \, .
\label{eq:tau-to-N}
\end{equation}
We observe that, at variance with the $N$-to-$\tau$ transform, this integral is
not analytic in $a$. In fact, for  negative $a$, it diverges for any $N$.
Furthermore, the logarithmic expansion of Eq.~(\ref{eq:tau-to-N}) contains factorially
growing terms. In fact, setting $x=N\tau$ and $\lambda_N=a \log N$, we can write
\begin{eqnarray*}
{\cal R}_N
& = &
\frac{1}{N}
\int_0^{\infty} \mathd x \,
\exp \left[ - a \ln^2 \frac{x}{N}  \right]\, e^{- x}
\\
& = &
\frac{\exp \left[ - \frac{\lambda_N^2}{a}  \right]}{N}
\int_0^{\infty} \mathd x \,
\exp [- a \ln^2 x + 2\lambda_N \ln x] \, e^{- x}
\\
& = &
\frac{\exp \left[ - \frac{\lambda_N^2}{a}  \right]}{N}
\int_0^{\infty} \mathd x \,
e^{- x} \, x^{2\lambda_N} \,
\sum_{m=0}^\infty \frac{(- a)^m}{m!}
\ln^{2 m} x
\, .
\end{eqnarray*}
By performing the change of variables $z = x^{1 + 2\lambda_N}$, so that $\mathd z
= (1 + 2\lambda_N) \,x^{2\lambda_N}\, \mathd x$, we can see that the terms of the series,
\begin{equation}
\int_0^{\infty}
\frac{\mathd z\, e^{-z^{\frac1{1+2\lambda_N}}}}{1+2\lambda_N}\,
\frac{(-a)^m}{m!} \,
\ln^{2m} z^{\frac1{1+2\lambda_N}}
\, = \,
\int_0^{\infty}
\frac{\mathd z \, e^{- z^{\frac1{1+2\lambda_N}}}}
{(1+2\lambda_N)^{2m+1}} \,
\frac{(- a)^m}{m!}
\ln^{2 m} z
\, ,
\end{equation}
exhibit a factorial growth that arises from the $z\to 0$ integration boundary, and is due to
the fact that
\begin{eqnarray*}
\int_0^1
\mathd z \,
\ln^{2 m} z 
& = &
(2 m) !
\,\, \approx \,\,
4^m \, (m!)^2.
\end{eqnarray*}
Thus we see that, as in the case of the resummation of threshold logarithms discussed in
Ref.~\cite{Catani:1996dj,Catani:1996yz}, violation of kinematic factorisation can lead
to a logarithmic expansion that, although correct from the point of view of logarithmic counting,
exhibits an undesired behaviour.

\section{The leading-logarithmic case}\label{sec:LL case}
The cumulative distribution at leading-logarithmic (LL) level is given by
\begin{equation}
\label{eq: N space cumulant LL}
\Rres(\tau)
\, = \,
\frac{1}{2 \pi i}
\int_\Cnt \frac{\mathd N}{N} \,
e^{N \tau}
e^{\frac{1}{a_s \beta_0} g_1(\lambda)}\, ,
\end{equation}
where $\lambda=a_s \beta_0 \log N$, 
\begin{equation}
\label{eq: g1}
g_1 (\lambda)
\, = \,
- \frac{A^{(1)}}{\beta_0 }
\Big[
(1 - 2 \lambda) \log (1 - 2 \lambda)
-
2 (1 - \lambda) \log (1 - \lambda)
\Big]
\, ,
\end{equation}
and $\beta_0 = (11 C_A - 2 N_f)/12$. Notice that
\begin{equation}
  g_1(\lambda)=-A^{(1)}\frac{\lambda^2}{\beta_0} +{\cal O}(\lambda^3)
\end{equation}
so that the leading term in the exponent of Eq.~(\ref{eq: N space cumulant LL}) is $-A^{(1)}a_s \log^2 N$, i.e.~the double-logarithmic term.

In line with the DL example of the previous section, we assume here that
Eq.~(\ref{eq: N space cumulant LL}) represents the full resummation result,
since our purpose in this section is only to study the subleading terms generated by inverting the conjugate space result to direct space.
We will transform Eq.~(\ref{eq: N space cumulant LL}) into a
direct-space result, and examine its structure in terms of direct-space
logarithmic counting.

%\begin{equation}
%a_s
%\, = \,
%\frac{\alpha_s(Q)}{\pi}
%\, ,
%\qquad\quad
%A^{(1)}
%\, = \,
%C_F
%\, ,
%\qquad\quad
%\beta_0
%\, = \,
%\frac{11 C_A - 2 N_f}{12}
%\, .
%\end{equation}

The function $g_1(\lambda)$ has branch points at the Landau poles $\lambda=1/2$, and $\lambda=1$. As a consequence, there is an ambiguity in the integration contour $\Cnt$, associated with the different possibilities to circle the branch points. It is typical to choose the $c$ constant in the H\"ankel contour in such a way that the Landau pole lies to the right of $\Cnt$. This is generally possible, since the closest Landau pole is at $N=\exp[1/(2a_s \beta_0)]\gg1$.
This procedure is known as ``minimal prescription'', see Refs.~\cite{Catani:1996dj,Catani:1996yz}. The choice of alternative prescriptions that differ by non-perturbative power corrections has a negligible impact, as we show for the case of the Borel prescription \cite{Forte:2006mi,Abbate:2007qv} in Appendix \ref{app:borel}.

It is convenient to define $x = N \tau$ so that
\beq
\lambda
\, = \,
a_s \beta_0 \log N
\, = \,
\lambda_x + \lambda_\tau
\, ,
\qquad\quad
\lambda_x
\, = \,
a_s \beta_0 \log x
\, ,
\qquad\quad
\lambda_{\tau}
\, = \,
a_s \beta_0 \log \frac{1}{\tau}
\, .
\eeq
The full cumulative distribution is written as
\beq
\Rres (\tau)
& = &
\frac{1}{2 \pi i}
\int_\Cnt \frac{\mathd x}{x} \,
e^x e^{\frac{1}{a_s \beta_0} g_1(\lambda)}
\nnb\\[4pt]
& = &
\frac{1}{2 \pi i}
\int_\Cnt \frac{\mathd x}{x} \,
e^x e^{\frac{1}{a_s \beta_0} g_1(\lambda_{\tau})
+ \frac{\lambda_x}{a_s \beta_0} g_1'(\lambda_{\tau}) + \frac{1}{a_s \beta_0} H_1 (\lambda_\tau, \lambda_x)}
\, ,
\eeq
having set $g_1^{(n)}(\lambda) = \mathd^n g_1(\lambda)/\mathd \lambda^n$, and
\beq
\label{eq:Hexpansion}
\frac{1}{a_s \beta_0}
H_1 (\lambda_\tau, \lambda_x)  
& = &
\frac{1}{a_s \beta_0}
\bigg(
g_1(\lambda_{\tau} + \lambda_x)
-
g_1(\lambda_{\tau})
-
\lambda_x \, g_1'(\lambda_\tau)
\bigg)
\nnb\\[4pt]
& = &
\frac{1}{a_s \beta_0}
\sum_{j=1}^\infty
\frac1{(j+1)!} \,
g_1^{(j+1)}(\lambda_\tau) \,
\lambda_x^{j+1}
\nnb\\[4pt]
& = &
\sum_{j=1}^\infty
\frac{(a_s \beta_0)^j}{(j+1)!} \,
g_1^{(j+1)}(\lambda_\tau) \,
\log^{j+1}x \label{eq: Expanded H}
%\nnb\\[4pt]
%& = &
%\frac{a_s\beta_0}{2} \,
%g_1''(\lambda_\tau) \,
%\log^2x
%+
%\dots
\, ,
\eeq
so that
\beq
\label{eq: R tau inverse Gamma}
\Rres (\tau)
& = &
e^{\frac{1}{a_s \beta_0} g_1 (\lambda_{\tau})}
\exp
\left(\frac{1}{a_s \beta_0}
H_1 \Big(\lambda_\tau, a_s \beta_0
\frac{\mathd}{\mathd g'_1(\lambda_{\tau})}\Big)
\right)
\frac{1}{2 \pi i}
\int_\Cnt
\frac{\mathd x}{x} \,
e^x e^{ \frac{\lambda_x}{a_s \beta_0}
g_1'(\lambda_{\tau})}
\nnb\\
& = &
e^{\frac{1}{a_s \beta_0} g_1 (\lambda_{\tau})}
\exp
\left(\frac{1}{a_s \beta_0}
H_1 \Big(\lambda_\tau, a_s \beta_0
\frac{\mathd}{\mathd g'_1(\lambda_{\tau})}\Big)
\right)
\frac{1}{\Gamma \big[1 - g_1' (\lambda_{\tau})\big]}
\, .
\eeq
\\
The differential distribution is given by
\beq
\frac{\mathd \Rres (\tau)}{\mathd \tau}
& = &
\frac{\mathd}{\mathd \tau}
\frac{1}{2 \pi i}
\int_\Cnt \frac{\mathd N}{N} e^{N \tau} e^{\frac{1}{a_s \beta_0} g_1(\lambda)}
%\exp (\log N f_1 (a_s b_0 \log N))
\, = \,
\frac{1}{2 \pi i}
\int_\Cnt \mathd N e^{N \tau}
e^{\frac{1}{a_s \beta_0} g_1(\lambda)}
%\exp (\log N f_1 (a_s b_0 \log N))
\, ,
\eeq
leading to
\beq
\frac{\mathd \Rres (\tau)}{\mathd \tau}
& = &
\frac{1}{2 \pi i}
\int_\Cnt \frac{\mathd x}{\tau} \,
e^x e^{\frac{1}{a_s \beta_0} g_1(\lambda_{\tau})
+ \frac{\lambda_x}{a_s \beta_0} g_1'(\lambda_{\tau})
+ \frac{1}{a_s \beta_0} H_1 (\lambda_\tau, \lambda_x)}
\nnb\\
& = &
\frac1\tau \,
e^{\frac{1}{a_s \beta_0} g_1 (\lambda_{\tau})}
\exp
\left(\frac{1}{a_s \beta_0}
H_1 \Big(\lambda_\tau, a_s \beta_0
\frac{\mathd}{\mathd g'_1(\lambda_{\tau})}\Big)
\right)
\frac{1}{2 \pi i}
\int_\Cnt
\mathd x \,
e^x e^{ \frac{\lambda_x}{a_s \beta_0}
g_1'(\lambda_{\tau})}
\nnb\\
& = & \label{eq:LL N form}
\frac1\tau \,
e^{\frac{1}{a_s \beta_0} g_1 (\lambda_{\tau})}
\exp
\left(\frac{1}{a_s \beta_0}
H_1 \Big(\lambda_\tau, a_s \beta_0
\frac{\mathd}{\mathd g'_1(\lambda_{\tau})}\Big)
\right)
\frac{1}{\Gamma \big[\!-g_1'(\lambda_{\tau})\big]}
\, .
\eeq
The prefactor in Eq.~\eqref{eq:LL N form} represents the leading-logarithmic
result in direct space. The exponent in round bracket
has an expansion in powers of $a_s$ starting at first order
(see Eq.~\eqref{eq: Expanded H}).
By expanding the exponential in powers of $a_s$ we obtain the subleading-logarithmic terms to all the desired orders.

Once we truncate Eq.~\eqref{eq:LL N form} to a definite order $n$,
there are several alternative
ways to organise the expression for the differential cross section, which
  disagree only by terms of order higher than $n$.
%distribution as a truncated expansion in $a_s$ at fixed $\lambda_\tau$. Defining
%\beq
%C_0
%\, = \,
%\frac{1}{\Gamma\big[\!-g_1'(\lambda_{\tau})\big]}
%\, ,
%\eeq
We can write
\beq
\label{eq:diff}
\frac{\mathd \RresA^{(\mathd,n)} (\tau)}{\mathd \tau}
& = &
\frac{1}{\tau} \,
e^{\frac{1}{a_s \beta_0} g_1 (\lambda_{\tau})}
\sum_{i = 0}^n C_i \, a_s^i 
\, ,
\\
\label{eq:diffexp}
\frac{\mathd \RresB^{(\mathd,n)} (\tau)}{\mathd \tau}
& = &
\frac{1}{\tau} \,
e^{\frac{1}{a_s \beta_0} g_1 (\lambda_{\tau})}
\exp \Big(\sum_{i = 0}^n C'_i \, a_s^i \Big)
\, ,
\\
\label{eq:cum}
\frac{\mathd \RresC^{(\mathd,n)} (\tau)}{\mathd \tau}
& = &
\frac{\mathd}{\mathd \tau}
\left[
e^{\frac{1}{a_s \beta_0} g_1 (\lambda_{\tau})}
\sum_{i = 0}^n D_i \, a_s^i
\right]
\, ,
\\
\label{eq:cumexp}
\frac{\mathd \RresD^{(\mathd,n)} (\tau)}{\mathd \tau}
& = &
\frac{\mathd}{\mathd \tau}
\left[
e^{\frac{1}{a_s \beta_0} g_1 (\lambda_{\tau})}
\exp \Big( \sum_{i = 0}^n D'_i \, a_s^i \Big)
\right]
\, ,
\eeq
with each prescription amounting to a slightly different
definition of the subleading orders. By keeping only the
prefactor we have the LL approximation; including the $n=0$
term we have the next-to-LL (NLL) result, with $n=1$ we have the next-to-NLL (NNLL) one, and so on.

Here we focus on the coefficients $D_i$ and $D'_i$ implicitly defined in Eq.~\eqref{eq:cum} and Eq.~\eqref{eq:cumexp}.
We could not find an analytic form for their asymptotic behaviour.
However, we were able to compute them numerically for $i$ as large as 60 for $D_i$, and 30 for $D'_i$. We find numerically the following behaviour:
\begin{equation}
\label{eq:LLasymptotics}
D_n, \, D'_n
\, \propto \,
\frac{\Lf(n) \, \pi^n}{(1-2\lambda_\tau)^n}
\times
P_n
\, ,
\end{equation}
where the ``log-factorial'' function $\Lf(n)$ is defined as \beq
\Lf(n) \, \equiv \, \prod_{j=2}^{n} \, \log(j) \, , \eeq and $P_n$ is
a periodic function that can be approximately represented by \beq P_n
\, = \, \cos\left(\frac{n\pi}{4}\right) \, .  \eeq
The geometric
factor $(1-2\lambda_\tau)^{-n}$ in Eq.~(\ref{eq:LLasymptotics}) is
easily understood from Eq.~(\ref{eq:Hexpansion}) and Eq.~\eqref{eq:
  g1}, where one can note that the $(n+1)^{\rm th}$ derivative of
$g_1(\lambda_\tau)$ is dominated by a $(1-2\lambda_\tau)^{-n}$ power.
Notice that
\begin{equation}
\frac{a_s^n(Q)}{(1-2\lambda_\tau)^{n}}
\,\, \approx \,\,
a_s^n(\tau Q)
\, ,
\end{equation}
which is an exact relation if we use the leading-order expression for the
running coupling.

\begin{figure}[htb]
\includegraphics[page=1,width=0.48\textwidth]{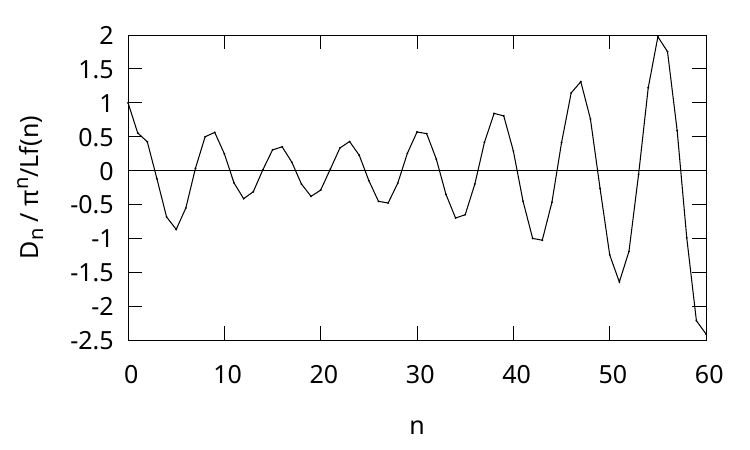}
\includegraphics[page=2,width=0.48\textwidth]{plotcoefs-final}
\caption{The $D_n$ (left panel) and $D_n'$ (right panel) coefficients divided by $\Lf(n) \, \pi^n$, computed numerically.}
\label{fig:termsplot} 
\end{figure}

In Fig.~\ref{fig:termsplot} we plot the $D_n$ and $D'_n$ coefficients divided by $\Lf(n) \, \pi^n$. 
%\begin{figure}[htb]
%  \includegraphics[page=18,width=0.48\textwidth]{plotcoefs-final}
%  \includegraphics[page=34,width=0.48\textwidth]{plotcoefs-final}
%  \includegraphics[page=36,width=0.48\textwidth]{plotcoefs-final}
%\caption{The $D'_n$ coefficients, computed numerically, multiplied by $(1-2\lambda_\tau)^n/(\Lf(n) \, \pi^n)$, for different values of $\lambda_\tau$. {\color{red} Reduce to a single figure, independent of $\lambda$} (PaoloT)}
%\label{fig:exptermsplot} 
%\end{figure}
The plots show the quality of the approximate representation
of Eq.~(\ref{eq:LLasymptotics}).

The asymptotic behaviour of formula~\eqref{eq:LLasymptotics} leads to an expansion with
zero radius of convergence, i.e.~to an asymptotic series.  By virtue
of the oscillating behaviour of the $P_n$ factor, the series can be resummed without
ambiguities.  The fact remains, however, that by truncating the
series at some fixed order, one neglects terms that are not necessarily small, and
the result may depend upon the truncation order.  We also observe that
the growth of the coefficients is very weak, quite far from the
factorial growth that accompanies power corrections. In
Appendix~\ref{sec:resummationAmbiguities} we show in fact that the
ambiguity corresponding to the size of the minimal term of the series
amounts to an exponential suppression in $Q$.

It  should  be  remarked  that  the behaviour  just  outlined  in  the
direct-space expansion  is intimately related  to the presence  of the
Landau pole.
% In fact, if the expansion were convergent, there would be no
% ambiguity in the integration of the full formula stemming from the
% choice of the contour relative to the position of the Landau pole.
% In fact, if the expansion were convergent, one could integrate the
% full formula with no ambiguities related to the how the contour
% $\Cnt$ circles the Landau pole.
In fact, if we consider the LL Sudakov exponent
$g_1(\lambda)$ in Eq.~\eqref{eq: g1} as a function of the complex
variable $\log N$, the presence of a Landau singularity at
$\log N=1/(2 \,a_s\beta_0)$ causes the Taylor expansion
\eqref{eq:Hexpansion} of $g_1(\lambda)$ around $\log N=\log\frac1\tau$
to have a finite radius of convergence
$\rho_N = |\log\frac1\tau-1/(2 \,a_s\beta_0)|$ (and similarly for the
expansion around any other point).  When one integrates the terms of
such a power series expansion along the prescribed H\"ankel contour,
as done for instance in \eqref{eq: R tau inverse Gamma}, part of the
contour necessarily lies outside of the region of convergence, where
the power expansion ceases to be valid. The $\Rres(\tau)$ distribution
at LL can thus be written as a series of derivatives of the inverse
$\Gamma[1-g_1']$ function only in an asymptotic sense, namely for
$a_s\to 0$, in which limit indeed $\rho_N\to\infty$.

Some comments are in order. First, the behaviour of the expansion in
the LL case is to be contrasted with the DL case, for which we have
found excellent convergence properties in Sec.~\ref{sec:DL case}. At
DL the Sudakov exponent is indeed just a monomial in $\log N$, namely
an entire function of $\log N$. As such, the radius of convergence of
its expansion around $\log N=\log\frac1\tau$ (or around any other
expansion point) is $\rho_N\to\infty$, whence the expansion is valid
for all of the points of the H\"ankel contour.
We have further confirmed this interpretation beyond DL level by expanding the LL Sudakov radiator $g_1(\lambda)$ at a finite power in $a_s$, as well as considering a fixed-coupling version of the N$^4$LL (see next sections) Sudakov radiator, obtained by turning off running effects, $\beta_i\to0$. Both approximations yield a polynomial radiator in $\log N$, and we verify an optimal convergence of the $\Rres(\tau)$ expression to the conjugate-space formulation.\footnote{We note that employing an analytic QCD coupling \cite{Shirkov:1996cd,Shirkov:1997wi} in the $g_1$ function still leads to an asymptotic series, as in the LL case. Indeed, even with the analytic coupling, $g_1$ is not an entire function of $\ln N$.}

As a second comment, from the above analysis one may be tempted to conclude that the formulation in conjugate space is superior to the one in direct space, whenever the Sudakov exponent features a Landau pole. However this conclusion may be misleading.
In order to perform thrust resummation, conjugate-space formulae such as \eqref{eq: N space cumulant LL} are not the only possible starting point. Resummation can indeed be performed entirely in direct space, where the observable is defined. This is done, for example, in the CAESAR,
ARES~\cite{Banfi:2003je,Banfi:2001bz,Banfi:2014sua,Banfi:2018mcq}
approaches, based on the use of angular ordering to implement soft coherence~\cite{Marchesini:1987cf}.\footnote{These approaches will be further discussed in Sec.~\ref{sec:num-resummation}.} In the following we will return on the comparison of the direct and conjugate-space approaches
in several occasions.

\section{The full calculation up to N$^4$LL}\label{sec:NXLL}
The inclusion of higher-order logarithmic terms is achieved by defining
\begin{equation}\label{eq:gexpansion}
g (\lambda)
\, = \,
\frac{1}{a_s \beta_0} g_1 (\lambda_{})
+ f_2 (\lambda)
+ a_s \, f_3 (\lambda)
+ a_s^2 \, f_4 (\lambda)
+ a_s^3 \, f_5 (\lambda)
+ \dots
\, ,
\end{equation}
so that the cumulative distribution becomes
\beq
\label{eq:Rlogfull}
\Rres (\tau) & = & \frac{C(a_s)}{2 \pi i} \int_\Cnt \frac{\mathd x}{x} \, e^x
e^{g_{} (\lambda)}
\nnb\\
& = & \frac{C(a_s)}{2 \pi i} \, e^{g_{}(\lambda_{\tau})} \int_\Cnt \frac{\mathd
  x}{x} \, e^x e^{\frac{\lambda_x}{a_s\beta_0} g_1' (\lambda_\tau) + H
  (\lambda_{\tau}, \lambda_x) }
\nnb\\
& = & e^{g_{}(\lambda_{\tau})} \, C(a_s) \exp \left( H \Big( \lambda_{\tau}, a_s
  \beta_0 \frac{\mathd}{\mathd g_1' (\lambda_\tau)}\Big) \right) \frac{1}{2 \pi
  i} \int_\Cnt \frac{\mathd x}{x} \, e^x e^{\frac{\lambda_x}{a_s\beta_0} g_1'
  (\lambda_\tau) }
\nnb\\
& = & e^{g_{} (\lambda_{\tau})} \, C(a_s) \exp \left( H \Big(\lambda_{\tau}, a_s
  \beta_0 \frac{\mathd}{\mathd g_1' (\lambda_\tau)} \Big) \right)
\frac{1}{\Gamma \big[1 - g_1' (\lambda_\tau)\big]} \, , \eeq where \beq H
(\lambda_{\tau}, \lambda_x) & = & g (\lambda_{\tau} + \lambda_x) - g
(\lambda_{\tau}) - \frac{\lambda_x}{a_s \beta_0} g_1' (\lambda_\tau) \, . \eeq

All the ingredients to achieve N$^{4}$LL resummation are
available~\cite{Becher:2008cf,Bruser:2018rad,Kelley:2011ng,Chen:2022yre,Baranowski:2021gxe,
  Baranowski:2022khd,Moult:2022xzt,Duhr:2022yyp,Duhr:2022cob,Baranowski:2024ysi,Baranowski:2024vxg,Moch:2004pa,
  Henn:2019swt,vonManteuffel:2020vjv,Herzog:2018kwj,Baikov:2009bg,Lee:2010cga,Gehrmann:2010ue,Lee:2022nhh}.
The relevant expressions for the resummation coefficients, including the finite
hard-virtual corrections $C(a_s)$ and the $f_i$ functions, can be found in the
appendix of Ref.~\cite{Aglietti:2025jdj}.

It is useful to explicitly report the first contributing terms to $H(\lambda_\tau,\lambda_x)$:
\beq
H(\lambda_\tau,\lambda_x)
& = &
a_s \beta_0 \log x \,
\Big(
f_2'(\lambda_\tau) +
a_s \, f_3'(\lambda_\tau) +
a_s^2 \, f_4'(\lambda_\tau) +
a_s^3 \, f_5'(\lambda_\tau) +
\dots
\Big)
\nnb\\
&&
+ \, 
\frac12
a_s \beta_0 \log^2 x \,
g_1''(\lambda_\tau) +
\dots
\, .
\eeq
%\beq
%  g_1 & : & \frac{1}{2} \frac{\lambda^2_x}{\beta_0 a} g_1'' (\lambda_x) =
%  \frac{a \beta_0}{2} \log^2 (x) g_1'' (\lambda_x)\\
%  g_2 & : & a^{} \beta_0 \log (x) f_2'\\
%  g_3 & : & a^2 \beta_0 \log (x) f_3'\\
%  g_4 & : & a^3 \beta_0 \log (x) f_4'\\
%  g_5 & : & a^4 \beta_0 \log (x) f_5'
%\eeq
If one wants to reach the maximum accuracy, i.e.~N$^4$LL, the expansion
of $H$ must be performed at order $a_s^3$, i.e.~it involves $f_4$ but not
$f_5$. In the prefactor $\exp(g (\lambda_\tau))$ we also include $f_5$, which yields a
term of order $a_s^3$.

We perform the computation of the differential distribution in two ways, i.e.
\beq
\frac{\mathd \Rres(\tau)}{\mathd \tau}
& = &
\label{eq:directdiff}
\frac{C(a_s)}{2\pi i}
\int_\Cnt
\frac{\mathd x}{\tau} \,
e^x e^{g(\lambda)}
\, ,
\\
\frac{\mathd \Rres(\tau)}{\mathd \tau}
& = &
\label{eq:diffcum}
\frac{C(a_s)}{2\pi i} \frac{\mathd}{\mathd \tau} \int_\Cnt
\frac{\mathd x}{x} \, e^x e^{g(\lambda)} \, , \eeq where in both cases
the integral over the H\"ankel contour is evaluated numerically using adaptive
Gaussian integration. In the second case, the derivative
with respect to $\tau$ is computed numerically from tabulated values of the corresponding integral. We have checked that the two results do
agree with one another to better than $5\times 10^{-5}$ over the whole
$\tau$ range. The two procedures are then also adopted in the
truncated versions of the same formulae:
\beq
\label{eq:directdifftrunc} 
\frac{\mathd \RresA^{(\mathd,n)}(\tau)}{\mathd \tau}
& = &
C(a_s) \,
\frac{1}{\tau} \,
e^{g(\lambda_\tau)} \,
\sum_{i=0}^n c_i \, a_s^i
\, ,
\\
\label{eq:directdifftruncexp}
\frac{\mathd \RresB^{(\mathd,n)}(\tau)}{\mathd \tau}
& = &
C(a_s) \,
\frac{1}{\tau} \,
e^{g(\lambda_\tau)} \,
\exp \Big( \sum_{i=0}^n c'_i \, a_s^i \Big)
\, ,
\\
\label{eq:directcumtrunc}
\frac{\mathd \RresC^{(\mathd,n)}(\tau)}{\mathd \tau}
& = &
C(a_s) \,
\frac{\mathd}{\mathd \tau}
\left[ e^{g(\lambda_\tau)} \,
\sum_{i=0}^n d_i \, a_s^i \right]
\, ,
\\
\label{eq:directcumtruncexp}
\frac{\mathd \RresD^{(\mathd,n)}(\tau)}{\mathd \tau}
& = &
C(a_s) \,
\frac{\mathd}{\mathd \tau}
\left[ e^{g(\lambda_\tau)} \,
\exp \Big( \sum_{i=0}^n d'_i \, a_s^i \Big) \right]
\, .
\eeq
% Table numbers from files cumplots-cum-N[1-4]LL-118.dat
The coefficients $c_i$, $c'_i$, $d_i$ and $d'_i$ are computed with the same computer-algebra technique used in the LL case. Results at N$^k$LL accuracy, with $k=1,2,3,4$, have been obtained considering the resummation functions in Eq.~\eqref{eq:gexpansion}
up to $f_{k+1}(\lambda)$, as well as setting $n=k-1$ as an upper bound for the sums in Eqs.~\eqref{eq:directdifftrunc} -- \eqref{eq:directcumtruncexp}.

\subsection{Comparison between direct and conjugate space up N$^4$LL}
\begin{figure}[htb]
\includegraphics[page=1,width=0.48\textwidth]{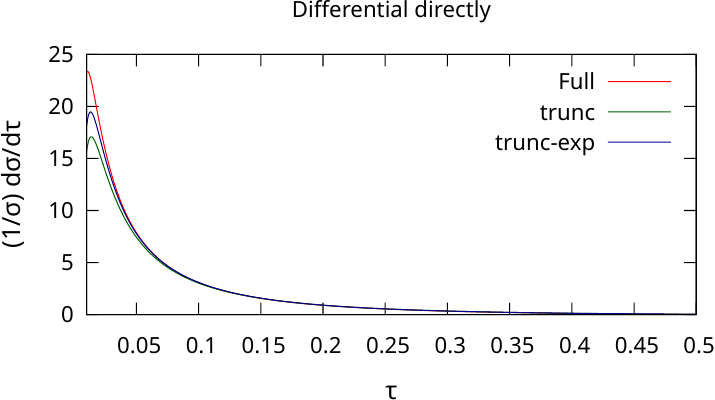}
\,\,
\includegraphics[page=2,width=0.48\textwidth]{checkConvNXLL-crop}
\,\,
\includegraphics[page=4,width=0.48\textwidth]{checkConvNXLL-crop}
\,\,
\includegraphics[page=5,width=0.48\textwidth]{checkConvNXLL-crop}
\caption{Differential thrust spectrum at N$^4$LL accuracy.
Left panels: the `Full' label refers to $\mathd \Rres(\tau)/\mathd \tau$ of Eq.~\eqref{eq:directdiff} (or equivalently \eqref{eq:diffcum}), while the `trunc' and `trunc-exp' labels refer to the expressions for $\mathd \RresA^{({\rm d},n)}(\tau)/\mathd \tau$ and $\mathd \RresB^{({\rm d},n)}(\tau)/\mathd \tau$ in Eqs.~(\ref{eq:directdifftrunc}), (\ref{eq:directdifftruncexp}). Right panels: `Full' corresponds again to Eq.~(\ref{eq:directdiff}), while the `trunc' and `trunc-exp' labels refer to the expressions for $\mathd \RresC^{({\rm d},n)}(\tau)/\mathd \tau$ and $\mathd \RresD^{({\rm d},n)}(\tau)/\mathd \tau$ in Eqs.~(\ref{eq:directcumtrunc}), (\ref{eq:directcumtruncexp}).
The lower panels display the ratios of the various direct-space predictions to the `Full' result.}
\label{fig:FullTruncTruncExp}
\end{figure}

\begin{figure}[htb]
\includegraphics[page=10,width=0.48\textwidth]{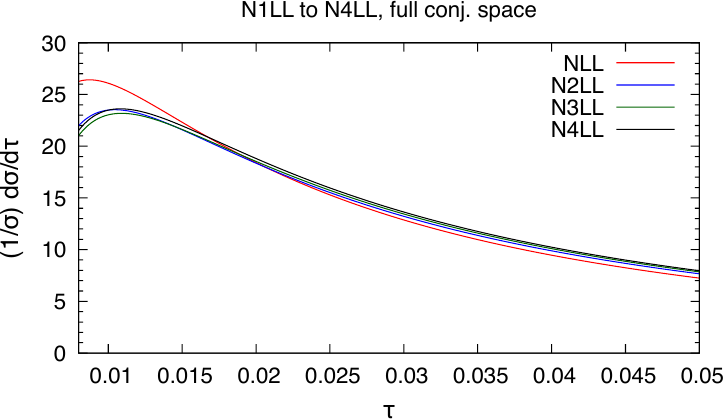}
\,\,
\includegraphics[page=11,width=0.48\textwidth]{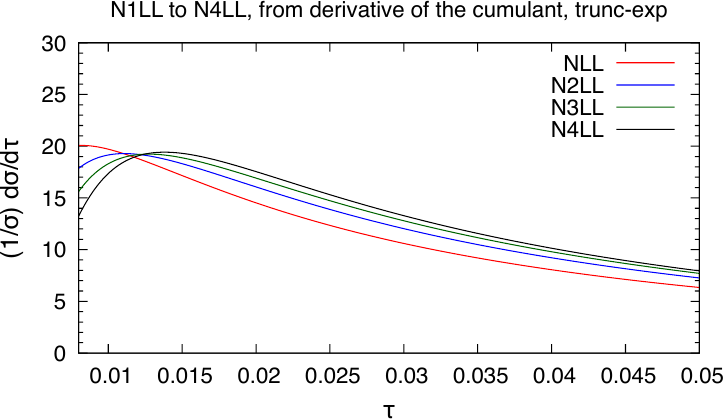}
\,\,
\includegraphics[page=12,width=0.48\textwidth]{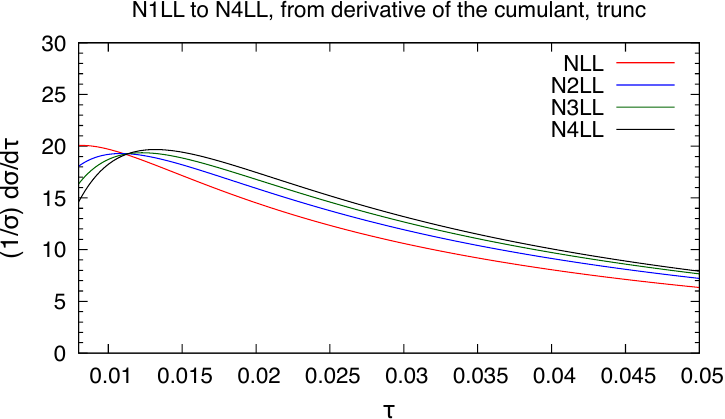}
\,\,
\includegraphics[page=13,width=0.48\textwidth]{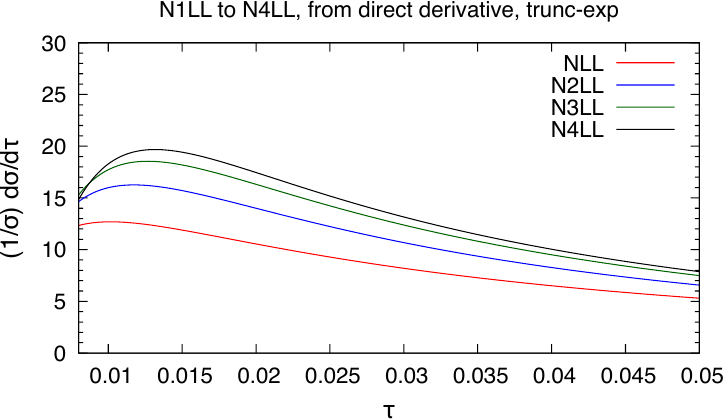}
\caption{Differential thrust spectrum obtained as a progression from NLL up to N$^4$LL. The upper left panel shows the full conjugate space result $\mathd \Rres(\tau)/\mathd \tau$ of Eq.~(\ref{eq:directdiff}); the upper right panel shows $\mathd \RresD^{({\rm d},n)}(\tau)/\mathd \tau$ of Eq.~(\ref{eq:directcumtruncexp}); The lower left panel displays $\mathd \RresC^{({\rm d},n)}(\tau)/\mathd \tau$, see Eq.~(\ref{eq:directcumtrunc}); finally, the lower right panel contains $\mathd \RresB^{({\rm d},n)}(\tau)/\mathd \tau$, see Eq.~(\ref{eq:directdifftruncexp}).}
\label{fig:NKKN4LL}
\end{figure}

In Fig.~\ref{fig:FullTruncTruncExp}
we display the result of the calculation for the N$^4$LL level of accuracy.
In Fig.~\ref{fig:NKKN4LL} we display the perturbative progression from NLL up to N$^4$LL, for the full result as well as for the various expansions in $\tau$ space that we have considered.
From the figures, we observe that in the peak region of the thrust distribution the direct- and conjugate-space formulations of the resummation yield different results. This difference remains significant even at the highest available logarithmic accuracy. Moreover, comparing the full conjugate-space result shown in the upper-left panel of Fig.~\ref{fig:NKKN4LL} with the direct-space results displayed in the remaining three panels, we find that the discrepancy decreases only very slowly as the perturbative order is increased.

By increasing the perturbative order one is increasing the number of logarithmic towers that are correctly predicted by the two formulations. However, it turns out that the subleading towers, which by construction are out of theoretical control in either formulation, are increasingly different with the order. This effect builds up an overall discrepancy between the two formulations which is roughly independent of the perturbative accuracy (at least, up to N$^4$LL) in the peak region.

We can get some insight on the reason for this behaviour by considering again the source of the discrepancy, namely the fact that the presence of a Landau pole in conjugate space casts a finite convergence radius for the Taylor expansion of the Sudakov exponent. Since the expansion point is $N=1/\tau$, the radius of convergence $|1/\tau-\exp(1/(2\,a_s\beta_0))|$ is very large at $\tau\sim {\cal O}(1)$, while it shrinks down to 0 as $\tau$ decreases and $1/\tau$ approaches the Landau pole. For large values of $\tau$ one then expects the bulk of the H\"ankel contour to lie within the region of convergence of the Taylor series, and in turn the two resummation formulations to be numerically similar. Conversely, at small $\tau$, most of the H\"ankel contour lies outside of the convergence region, leading to larger numerical discrepancies.

The size of the discrepancy is thus  driven by
the position of the Landau pole and of the expansion point $N=1/\tau$. As soon as one includes higher orders, the sensitivity of the resummed expression to the various regions in $N$ space (i.e.~to the various segments of the H\"ankel contour) tends to stabilise, and in turn the dependence of the discrepancy upon the logarithmic accuracy is not expected to have a clear perturbative convergence. In a theory with a smaller (larger) value of $\as\beta_0$ one would expect the discrepancy to kick in at smaller (larger) values of $\tau$, since the Landau pole would simply be shifted, on average, further from (closer to) the expansion point.
%We provide a further quantitative analysis of the difference between the two formulations in appendix \ref{sec:Nvstau}.

\subsection{The total cross section}
Resummation implies adding to differential cross sections an infinite
tower of corrections that are enhanced in specific phase-space
regions. When considering the integral of the differential cross
section, i.e.~the total cross section, the resummed formula still
consists of an infinite tower of perturbative corrections. However
these are not predictive, as the coefficients of the expansion do not
carry any enhancement.

Our resummation formula for the cumulative distribution
is normalised to the Born cross section (i.e.~it yields 1 if we set $\tau=1$ and $\as=0$), and thus it differs from 1 for physical values of $\as$. 
It is thus useful to examine whether the total cumulative distribution predicted by our resummation formula is well behaved, in the sense that the coefficients of its perturbative expansion are comparable in size to those obtained in the full theory, up to the orders at which they have been
evaluated.

The resummed formula for the total cumulative distribution
$\Rres(1)$ can be directly obtained from the cumulative distribution
by setting $\tau=1$, and thus $\lambda_\tau=0$. For instance, in the LL approximation, if we take the direct-space expression $\RresC^{(\mathd,n)} (\tau)$ of 
Eq.~\eqref{eq:cum}, and set $\as=0.118$, we get for the expansion
\begin{equation}
\label{eq:sigmatotLL}
\RresC^{(\mathd,n)} (1)
\, = \,
1+
0.6569 \cdot 10^{-1} +
0.4104 \cdot 10^{-2} -
0.1510 \cdot 10^{-3} +
{\cal O}(\as^4)
\, .
\end{equation}
Thus, nothing ill behaved takes place at this level.
Notice that by summing all terms of Eq.~\eqref{eq:sigmatotLL} we obtain 1.06965, while the full resummation yields 1.06947,
meaning that missing terms introduced by the resummed formula beyond N$^4$LL yield a contribution $1.8\times10^{-4}$ to the
QCD correction to the total cross section, that is negligible for our purposes.

The asymptotic behaviour of the coefficients is illustrated in the left panel of  Fig.~\ref{fig:termsplot}. Also in
this case a log-factorial growth with oscillating coefficients is visible, but it is associated with very
tiny effects.

In table~\ref{tab:sigtotal}
\begin{table}[htb]
  \begin{center}
    \begin{tabular}{|c|c|c|c|}
      \hline
      Order & $\Rres(1)$ Eq.~(\ref{eq:diffcum}) & $\RresC^{({\rm d},n)}(1)$  Eq.~(\ref{eq:directcumtrunc})
            & $\RresD^{({\rm d},n)}(1)$ Eq.~(\ref{eq:directcumtruncexp}) \\
      \hline
      NLL     & 1.0685  &  1  & 1 \\
      \hline
      NNLL & 1.0690 & 1.0436  & 1.0452 \\
      \hline
      N$^3$LL & 1.0743 & 1.0652  & 1.0665 \\
      \hline
      N$^4$LL & 1.0762 & 1.0727  & 1.0735  \\
      \hline
    \end{tabular}
  \end{center}
  \caption{\label{tab:sigtotal}   Total cumulative distribution as obtained from the conjugate and the direct-space formalisms, for $\as(\Mz)=0.118$.}
\end{table}
we display the total cumulative distribution as computed from the full N$^4$LL conjugate-space formula $\Rres(1)$ in Eq.~(\ref{eq:diffcum}) (or, equivalently, in the first line of Eq.~(\ref{eq:Rlogfull})), compared with the values obtained from the direct-space formulae $\RresC^{({\rm d},n)}(1)$ and $\RresD^{({\rm d},n)}(1)$ in Eqs.~(\ref{eq:directcumtrunc}) and (\ref{eq:directcumtruncexp}).
\section{Missing higher-order uncertainties and matching}
\label{sec:scalevar-matching}

It is important to establish to what extent traditional scale-variation bands,
routinely employed to estimate uncertainties stemming from missing higher
orders, cover the differences between the two resummation approaches. To this
end, we present scale-variation bands for both conjugate-space resummation
(labelled as N$^k$LL$_N$) and direct-space resummation (N$^k$LL$_\tau$). We now
choose as our default for the direct-space result the $\RresD^{(\mathd,n)}(\tau)$ variant defined in
Eq.~\eqref{eq:directcumtruncexp}.
This choice is motivated by the fact that the derivative of the cumulant, shown in the upper-right and lower-left panels of Fig.~\ref{fig:NKKN4LL}, exhibits a more stable behaviour as the perturbative order is increased than the direct derivative formula displayed in the lower-right panel. We have chosen to use the formula with the truncated expression in the exponent, as this is the convention most commonly adopted in the literature.

\begin{figure}[htb]
  \includegraphics[width=0.48\textwidth]{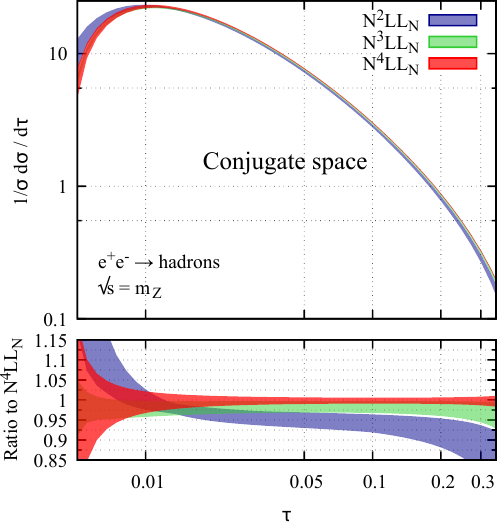}\hskip 0.3cm
  \includegraphics[width=0.48\textwidth]{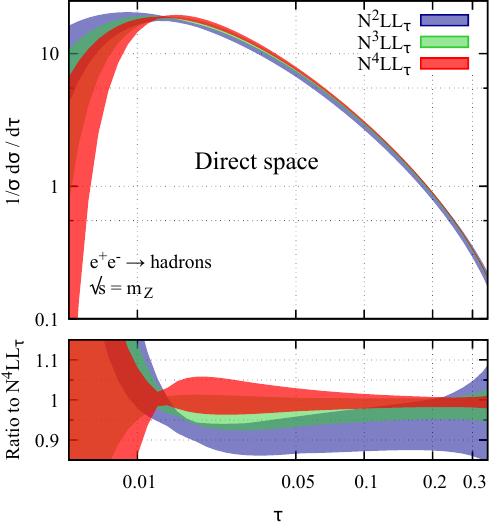}
  \vskip 0.3cm
  \begin{center}
    \includegraphics[width=0.48\textwidth]{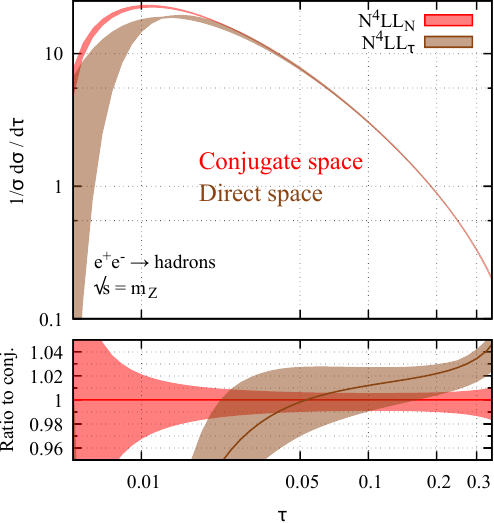}
  \end{center}
  \caption{Scale-variation bands for the conjugate and direct-space
    differential thrust distribution. The upper plots show results
    for the conjugate (left) and direct (right) spaces at
    increasing logarithmic orders, with uncertainties
    estimated through scale variations as detailed in the text.
    The ratio of the N$^4$LL$_N$ (formula~\eqref{eq:directdiff}) and  N$^4$LL$_\tau$
    (formula~\eqref{eq:directcumtruncexp}) results relative to
    the central value of the N$^4$LL$_N$ result is displayed in the lower plot.
    \label{fig:scalevar}}
\end{figure}

Our scale variation prescription amounts to varying independently the renormalisation scale $\mur$ and the resummation scale $\muL$
by a factor of two around their central values $\mur=\muL=Q$, excluding the scale combinations that yield $\muL/\mur=4$ or $1/4$. The
resummation scale $\muL$ is introduced by splitting the resummed logarithms as
\begin{align}
L_{\tau}
\, \equiv \,
\ln\frac{1}{\tau} & \,
\to \,
\ln\frac{\muL} {Q \tau} +
\ln \frac{Q}{\muL}
\, \equiv \,
L_{\tau;\muL} +
\ln \frac{Q}{\muL}
\, ,
\\
L_{N}
\, \equiv \,
\ln{N} & \,
\to \,
\ln\frac{\muL N }{Q} +
\ln\frac{Q}{\muL}
\, \equiv \,
L_{N;\muL} +
\ln\frac{Q}{\muL}
\, .
\end{align}
The results are displayed in Fig.~\ref{fig:scalevar}.
From the figure we see that the conjugate-space result has smaller scale-variation bands, which however overlap only at very small $\tau$ values. Above
the peak region, the series shows a slow convergence, with marginal overlap
between the N$^{3}$LL$_{N}$ and N$^{4}$LL$_{N}$ results. The direct-space result
shows the opposite trend, with a better convergence at moderate-large $\tau$
values, though within larger bands, and a worse convergence below the peak.

By comparing the two approaches at the N$^4$LL level, we see that the large
difference in the peak region is not covered by scale variations.
The two results overlap only between $\tau=0.02$ and $\tau=0.15$, while above this range the central predictions differ by a few percent.
These findings are consistent with our previous observation of relatively large differences beyond N$^4$LL between the two approaches. 

So far we have considered resummed predictions alone. In the region of moderate and large $\tau$ values, a matching to the fixed-order result is mandatory to obtain a reliable perturbative description of the spectrum. In
the following, we extend our analysis considering an additive matching
procedure, schematically defined as
\begin{equation}
\mathcal{R}_{T}^{{\rm N}^{k}{\rm LL} + {\rm N}^{k-1}{\rm LO}}(\tau)
\, = \,
\Rres^{{\rm N}^{k}{\rm LL} }(\tau)
\, + \,
\mathcal{R}_{T}^{{\rm N}^{k-1}{\rm LO}}(\tau)
\, - \,
\Rres^{{\rm N}^{k}{\rm LL} }(\tau)\big|_{\as^{k-1}}
\, ,
\end{equation}
where the rightmost term corresponds to the fixed-order expansion of the resummed
result up to ${\cal O}(\as^{k-1})$. The matching procedure requires one to turn off the resummed logarithms at large (small) values of $\tau$ ($N$). A possible way to achieve
this is by introducing modified logarithms to ensure the condition
$\mathcal{R}_{T}^{{\rm N}^{k}{\rm LL} + {\rm N}^{k-1}{\rm LO}} (\tau_{\rm max})= 1$
at the kinematical endpoint $\tau_{\rm max}$ of the cumulative distribution.
This amounts to applying the replacements
\begin{align}
\label{eq:modlogtau}
L_{\tau;\muL} & \to \,
\frac{1}{p}
\ln\left(
\left(\frac{\muL}{Q\tau}\right)^p+ 1 -\frac{1}{\tau^{\,p}_{{\rm max}}}\right)\;\Theta\left(\tau_{\rm max}- \frac{Q\tau}{\muL}\right)
\, ,
\\
\label{eq:modlogN}
L_{N;\muL} & \to \,
\frac{1}{p}
\ln\left(
\left(\frac{\muL N}{Q}\right)^p- N_{0}
\right)
\, , 
\end{align}
which induce subleading power corrections (scaling as $\tau^p$ or as $1/N^p$, respectively) in the results.
In the above equations, $\tau_{\rm max}$ is computed at any given perturbative
order, as reported for example in Ref.~\cite{Aglietti:2025jdj,Cacciari:2025xoo} , while the (negative) shift $N_{0}$ is determined by
imposing the unitary constraint
$\mathcal{R}_{T}^{{\rm N}^{k}{\rm LL} + {\rm N}^{k-1}{\rm LO}}(\tau_{\rm max}) = 1$.
In the following results, we use $p=1$ to facilitate a comparison with the results of Ref.~\cite{Aglietti:2025jdj} (with this choice, we find $N_0 \sim -1$). 
The fixed-order predictions up to order $\alpha_s^3$ are obtained as follows.
The order $\alpha_s$ result is fully analytic and has been known for a long time~\cite{DeRujula:1978vmq}. 
We extract the  $\alpha_s^2$ coefficient 
from the numerical tables of Ref.~\cite{Weinzierl:2009ms}, whereas
we use the \texttt{EERAD} code~\cite{Gehrmann-DeRidder:2014hxk} to compute the 
prediction at order $\alpha_s^3$.

\begin{figure}[t]
\centering
\includegraphics[width=0.48\textwidth]{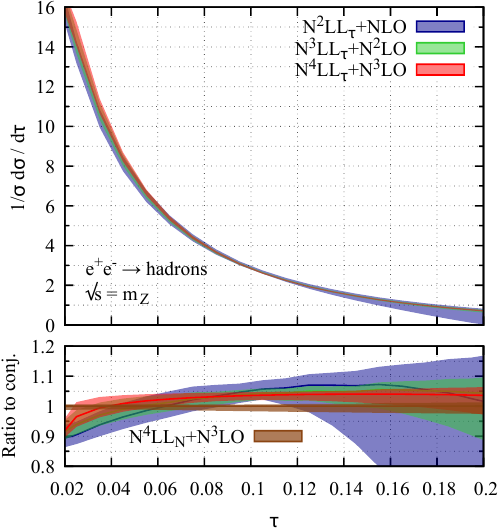}
\caption{Scale-variation bands for the matched thrust distribution.
The main panel shows the perturbative progression of the direct-space results, including uncertainty bands as detailed in the text.
The lower inset shows the ratio of the direct-space results, as well as of the conjugate-space N$^4$LL$_N$+N$^3$LO band, relative to the central value of the N$^4$LL$_N$+N$^3$LO prediction.
\label{fig:scalevar-matched}}
\end{figure}

In Fig.~\ref{fig:scalevar-matched}, we show the matched results for the
direct-space resummation at different perturbative orders (main panel) and their
ratio (lower inset) to the conjugate-space result at N$^{4}$LL+N$^{3}$LO. We
focus on the region above the peak of the distribution, where the matching is
more relevant. We observe that the difference between the two approaches
decreases upon including higher perturbative orders, and is below 5\% at
N$^{4}$LL+N$^{3}$LO. Although the uncertainty bands at this accuracy do not
cover this difference in the region $\tau \in [0.06,0.15]$, which has been
recently used for fitting the strong coupling~\cite{Benitez:2024nav}, we note
that the few-percent difference between these predictions is considerably
smaller than the one shown in Ref.~\cite{Aglietti:2025jdj}, where it reaches
$20\%$ at $\tau\sim 0.15$ with a much steeper slope.\footnote{A detailed
  comparison with the results of Ref.~\cite{Aglietti:2025jdj} is ongoing.}

The discrepancy between direct and conjugate results in the matching region suggests that
alternative matching prescriptions should be considered in order to obtain a more conservative and reliable uncertainty band.
In fact, neither $\muL$ nor $\muR$ are associated to a matching uncertainty.
In order to explore matching systematics, we consider for our direct-space predictions the alternative modified logarithms introduced in Ref.~\cite{vanBeekveld:2025zjh}.
We refer the reader to Ref.~\cite{vanBeekveld:2025zjh} for the exact definition of such a prescription.
Here it suffices to say that, at variance with the modified logarithms defined in Eq.~\eqref{eq:modlogtau}, those of Ref.~\cite{vanBeekveld:2025zjh} are designed to smoothly vanish at a matching scale $\muM$, which is independent of $\muL$.
The scale $\muM$ can be varied to probe the uncertainty due to the matching procedure, offering an additional free parameter besides $p$.
In the results shown below, we choose a central value $\muM/Q = 0.4$, and we vary it between $0.3$ and $0.5$ for all
the $\muL$, $\muR$ variations considered above. 
The final uncertainty is constructed as an envelope of the ensuing 21 variations.

\begin{figure}[t]
  \centering
  \includegraphics[width=0.48\textwidth]{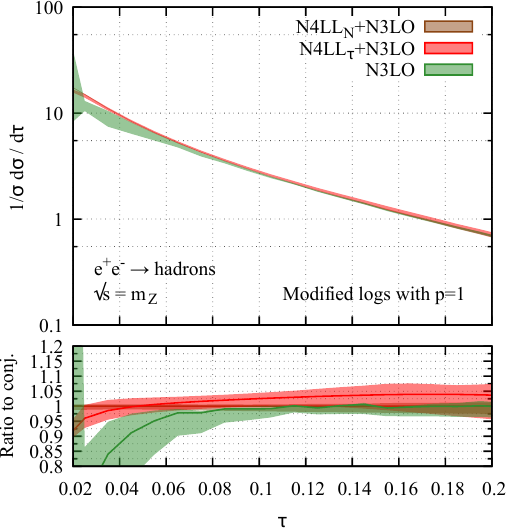}
  \includegraphics[width=0.48\textwidth]{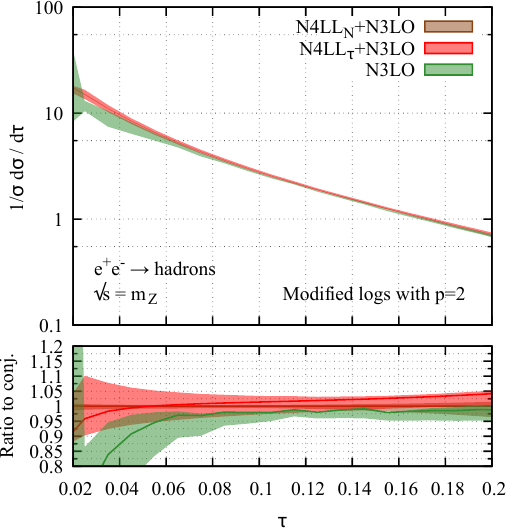}
  \caption{Scale-variation bands for the N$^3$LO fixed-order result, as well as for conjugate-space and direct-space matched results at N$^4$LL+N$^3$LO accuracy. The lower panel displays the three predictions normalised to the central value of the conjugate-space N$^4$LL$_N$+N$^3$LO result. On the left (right) we show distributions obtained with the modified logarithms of Eq.~\eqref{eq:modlogN} for conjugate space, and those of Ref.~\cite{vanBeekveld:2025zjh} for direct space, with $p=1$ ($p=2$).
    \label{fig:scalevar-matched-new}}
\end{figure}

The results are displayed in the left panel of Fig.~\ref{fig:scalevar-matched-new}, where we also show the fixed-order predictions at N$^3$LO.
The latter has been obtained with a central $\muR = Q \,\tau$ and variations of a factor of two around the central value.
The adoption of this scale in the fixed-order prediction is motivated by the observation that a dynamical scale should follow more closely the physical scale in each $\tau$ bin.
Such a choice improves the agreement between fixed order and resummation with respect to computations with a fixed scale $\muR = Q$.
A comparison with Fig.~\ref{fig:scalevar-matched} shows that the curve obtained with the new modified logarithms has a similar shape as that obtained with Eq.~\eqref{eq:modlogtau}, however it displays a more conservative uncertainty band, now overlapping with the conjugate-space result (which still uses Eq.~\eqref{eq:modlogN}).

We observe that modifying the power $p$ of the modified logarithms provides another way of estimating matching uncertainties, as it affects higher-order power corrections.
In the right panel of Fig.~\ref{fig:scalevar-matched-new} we show the results obtained setting $p=2$.
By comparing the results to those on the left panel, we observe a similar level of agreement between the predictions in direct and in conjugate space.
Direct-space predictions with $p=2$ feature a larger uncertainty band towards the peak of the distribution, while the band is slightly smaller above $\tau \sim 0.1$.

This exercise allows us to conclude that the difference between the results obtained in direct and conjugate spaces at N$^4$LL+N$^3$LO amounts to a few percent above the peak region, and can be covered using a sufficiently conservative uncertainty estimate.
In the matching region a reliable estimate of the uncertainty band is non trivial, and necessarily requires an account of matching systematics.
A precise assessment of theoretical uncertainties is crucial for determining the strong coupling constant, as larger uncertainties are expected to hinder the precision with which $\alpha_s$ can be extracted from fits to the thrust distribution.
Alternative prescriptions to estimate missing higher-order uncertainty may also be explored, for instance using theory nuisance parameters rather than scale variations \cite{Tackmann:2024kci}.
Due to the interplay with the modelling of non-perturbative corrections, which we do not consider in this work, we postpone this discussion to future studies.

\section{Approximations involved in the derivation of conjugate-space resummation}
\label{sec:shower-conjugate}
In this section, we derive the conjugate-space NLL resummation formula for thrust, starting from the factorisation of the multiple-radiation matrix element and phase space. In doing so, we analyse the numerical impact of the approximations involved to obtain expressions such as \eqref{eq: N space cumulant LL}.

We start from the master formula for the NLL cumulative thrust
distribution,
\begin{align}
\label{eq:master-shower}
\Rres^{\rm NLL}(\tau) &
= \,
\sum_{m=0}^{\infty}
\frac{1}{m!}
\int\prod_{i=1}^{m} \,
[\rd k_{i}] \,
\big[ M^{2}(k_{i}) \big]_+ \,
\Theta
\Big(\tau - \sum_{i=1}^{m}\tau(k_{i})\Big)
\, ,
\end{align}
where $[\rd k_i]$ is the phase-space measure for a single gluon emission, and 
$M^{2}(k_i)$ is the corresponding squared matrix element. Virtual corrections to gluon radiation are automatically included by means of the `plus' distribution
\beq
\int [\rd k] \, \big[ M^2(k) \big]_+ \, f(k)
\, = \,
\int [\rd k] \, M^2(k) \, \big[ f(k) - f(0) \big]
\, .
\eeq
It is understood that the product with no terms (i.e. with $m=0$) is simply equal to 1.
In the measurement function $\Theta$ the expression for thrust in presence of $m$ emissions has been approximated as the sum of the individual contributions $\tau(k_i)$ from each emission, which is accurate up to power-suppressed effects. Furthermore, the complete matrix element for $m$ gluon radiations has been approximated as the factorised product of single-emission matrix elements, which is accurate at NLL.

In order to discuss the connection between the master formula in Eq.~\eqref{eq:master-shower} and the conjugate-space formulation, we begin by rewriting the measurement function using the Laplace integral representation of the Heaviside function,
\begin{equation}
\label{eq:LaplaceTheta}
\Theta\Big(
\tau-\sum_{i=1}^{m}\tau(k_i)
\Big)
\, = \,
\frac{1}{2\pi i}
\int_{\mathcal{C}}
\frac{\rd N}{N} \,
e^{N \tau} \,
\prod_{i=1}^{m} e^{-N \tau(k_i)}
\, ,
\end{equation}
where $\mathcal{C}$ denotes a contour parallel to the imaginary axis, lying to the right of all singularities of the integrand, in particular the rightmost one at $N=0$. Using Eq.~\eqref{eq:LaplaceTheta}, we can rewrite Eq.~\eqref{eq:master-shower} as
\begin{align}
\Rres^{\rm NLL}(\tau) & = \,
\frac{1}{2\pi i}
\int_{\mathcal{C}}
\frac{\rd N}{N} \,
e^{N \tau} \,
\sum_{m=0}^{\infty} \,
\frac{1}{m!}
\int \prod_{i=1}^{m} \,
[\rd k_{i}] \,
\big[ M^{2}(k_{i}) \big]_+ \,
e^{-N \tau(k_i)}
\notag \\
& = \,
\frac{1}{2\pi i}
\int_{\mathcal{C}}
\frac{\rd N}{N} \,
e^{N \tau} \,
\sum_{m=0}^{\infty} \,
\frac{1}{m!}
\left(
\int [\rd k] \,
\big[ M^{2}(k) \big]_+ \,
e^{-N \tau(k)}
\right)^m
\notag \\
& = \,
\frac{1}{2\pi i}
\int_{\mathcal{C}}
\frac{\rd N}{N} \,
e^{N \tau} \,
\exp\left[
\int [\rd k] \,
M^{2}(k) \,
\big(e^{-N \tau(k)}-1\big)
\right]
\, . \label{eq:Nspace}
\end{align}
In the last equality of Eq.~\eqref{eq:Nspace} we have made the action of the plus distribution explicit, in order to highlight the role of real and virtual contributions, corresponding to $e^{-N\tau(k)}$ and $-1$ respectively.
Finally, by introducing the NLL Sudakov radiator $\Rnll({\bar\tau})$ \cite{Banfi:2004yd} and its derivative,
\beq
\label{eq:RNLL dir space}
\Rnll({\bar\tau})
\, = \,
\int[\mathd k] \,
M^2(k) \,
\Theta(\tau(k)-{\bar\tau})
\, ,
\qquad \quad
\Rpnll({\bar\tau})
\, = \,
\frac{\rd \Rnll({\bar\tau})}{\rd \ln {1/\bar{\tau}}}
\, ,
\eeq
we obtain the following conjugate-space representation of the cumulative  cross section,
\begin{align}
\label{eq:Sig_full}
\Rres^{\rm NLL}(\tau)
\, = \,
\frac{1}{2\pi i}
\int_{\mathcal{C}}
\frac{\rd N}{N} \,
e^{N \tau}
\exp\left[
{\int_{0}^{1}
\frac{\rd \bar{\tau}}{\bar{\tau}} \,
\Rpnll(\bar{\tau}) \,
\big(e^{-N \bar{\tau}}-1\big)}
\right]
\, ,
\end{align}
which explicitly shows the complete analytic factorisation of the
resummation formula in Laplace space. Since no approximation affecting
subleading logarithms has been introduced, this expression is fully equivalent
to the original master formula \eqref{eq:master-shower} from the point of view of resummed perturbation theory.

In the context of conjugate-space resummation, it is customary to approximate the
integral in the exponent of Eq.~\eqref{eq:Sig_full} so as to allow its analytic evaluation while preserving the logarithmic accuracy of the resummation. Up to NLL accuracy, this amounts to the approximation \cite{Catani:1992ua}
\begin{equation}
\label{eq:theta_approx}
e^{-N \bar{\tau}}-1
\, = \,
- \,
\Theta\Big(
\bar{\tau} - \frac{1}{\overline{N}}
\Big)
\, + \,
{\cal O}({\rm NNLL})
\, ,
\end{equation}
with $\overline{N}=Ne^{\gamma_{E}}$. We refer to Eq.~\eqref{eq:theta_approx} as the {\it theta-function approximation}. From now on we will denote by $\RresfullNLL$ the cross section in Eq.~\eqref{eq:Sig_full}, where this approximation is not applied, and by $\Rres^{\rm NLL, trunc}$ the cross section obtained by means of Eq.~\eqref{eq:theta_approx}.

In the following, we present a numerical study of the impact of the
theta-function approximation in Eq.~(\ref{eq:theta_approx}). For ease of
presentation, we restrict ourselves to the LL level, however we have verified
that the same conclusions also hold at NLL. In order to regulate the Landau pole
appearing in the full thrust cumulative distribution \eqref{eq:Sig_full}, we
adopt the simple prescription to cut off the $\bar \tau$ integration at
$\tau_{\rm L} = e^{-\frac{1}{2 a_{s} \beta_0}} = \Lambda^{(\rm LL)}_{\rm QCD}/Q$.\footnote{We
  have verified that the results are insensitive to an upward variation of $\tau_{L}$ by
  a factor of two.}
% We then compare the shower prediction, computed according to
% \begin{align}
% \Rres^{\rm LL, shower}(\tau) &= \int_{\tau_{\rm LP}}^{1}\rd \left(\ln \frac{1}{\tau_{1}}\right)
% \frac{\rd}{\ln \frac{1}{\tau_{1}}}\left[e^{-\Rll(\tau_1)}\right] \notag \\ \times  \sum_{m=0}^{\infty}\frac{1}{m!}
% &\left[\prod_{i=2}^{m+1}
% \int_{\tau_{\rm LP}}^{\tau_{1}}
% \rd \left(\ln \frac{1}{\tau_{i}}\right)
% \frac{\rd}{\rd \ln \frac{1}{\tau_{i}}}
% \Delta_{\rm LL}(\tau_{i-1},\tau_{i})\right]
% \Delta_{\rm LL}(\tau_{m+1},\tau_{\rm LP})
% \Theta\!\left(\tau-\sum_{i=1}^{m+1}\tau_{i}\right)\,,
% \end{align}
The full LL spectrum thus reads
\begin{align}
\label{eq:Sig_full_LL}
\frac{\rd}{\rd \tau} \RresfullLL(\tau)
& = \,
\frac{1}{2\pi i}
\int_{1 - i\infty}^{1 + i\infty}
\rd N \,
e^{N \tau}
\exp
\left[{\int_{\tau_{\rm L}}^{1}
\frac{\rd \bar{\tau}}{\bar{\tau}} \,
\Rpll(\bar{\tau}) \,
\big(e^{-N \bar{\tau}}-1\big)}
\right]
\notag \\
& \equiv \,
\frac{e^{\tau}}{2\pi}
\int_{-\infty}^{+\infty} \rd x \,
e^{i x \tau} \,
e^{-\widetilde{R}_{\rm LL}
\big(N \, = \, 1+ix; \, \tau_{\rm L}\big)}
\, ,
\end{align}
where we have defined
\begin{equation}
\widetilde R(N; \, \tau_{\rm L})
\, = \,
-
{\int_{\tau_{\rm L}}^{1}
\frac{\rd \bar{\tau}}{\bar{\tau}} \,
\Rpll(\bar{\tau}) \,
\big(e^{-N \bar{\tau}}-1\big)}
\, .
\end{equation}
We stress that in Eq.~\eqref{eq:Sig_full_LL} the Landau pole is regularised by the introduction of the $\tau_{\rm L}$ cutoff. As a result, the integral in the
exponent can be evaluated numerically, and the inverse Laplace transform does
not require the use of the minimal prescription \cite{Catani:1996dj,Catani:1996yz}.

Conversely, making use of the theta-function approximation at LL we get
\begin{equation}
\label{eq:dSig_trunc_LL}
\frac{\rd }{\rd \tau}
\Rres^{\rm LL, trunc}(\tau)
\, = \,
\frac{1}{2\pi i}
\int_{\mathcal{C}} \rd N \,
e^{N \tau} \,
e^{-\Rll(1/N)}
\, ,
\end{equation}
where the derivative of the LL radiator reads
\begin{equation}
\Rpll(\bar\tau)
\, = \,
-\frac{2 A_{1}}{\beta_{0}}
\Big[
\ln(1-2\lambda_{\bar\tau})
-
\ln(1-\lambda_{\bar\tau})
\Big]
\, ,
\end{equation}
and $\lambda_{\bar\tau}=a_{s}\beta_{0} \ln(1/\bar\tau)$.
When using Eq.~\eqref{eq:theta_approx} the $\bar\tau$ integration is cut at $1/\overline N$ by construction, hence the $\tau_{\rm L}$ cutoff need not be introduced. This in turn gives the $N$ integrand in Eq.~\eqref{eq:dSig_trunc_LL} a
sensitivity to the Landau pole for values of $\ln N \sim 1/(2 a_{s} \beta_{0})$, which requires a prescription to perform the inverse Laplace transform.

We anticipate here that Eq.~(\ref{eq:master-shower}) can be implemented in direct space as a parton-shower algorithm ordered in the $\tau$ variable, as we will prove in Sec.~\ref{sec:num-resummation}. In the following comparisons, we
include the results of such a parton-shower implementation -- given by the LL version of Eq.~\eqref{eq:master-transfer-algo} --  in order to check its
numerical equivalence with Eq.~\eqref{eq:Sig_full_LL}.

\begin{figure}
\centering
\includegraphics[width=0.65\textwidth]{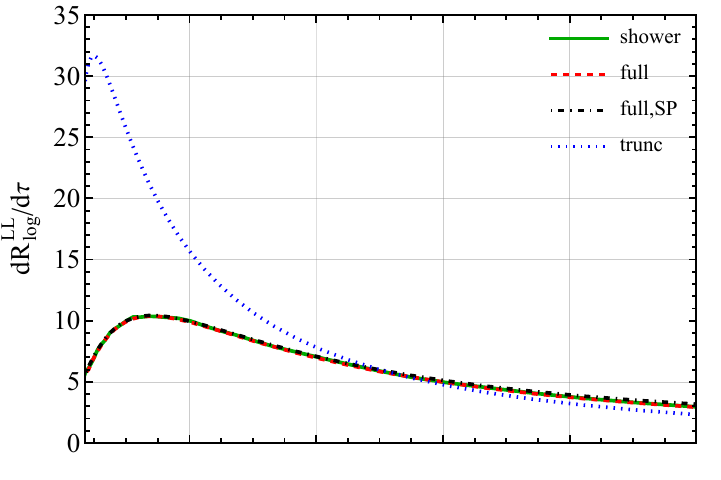}\
\vspace{-0.2cm}
\includegraphics[width=0.65\textwidth]{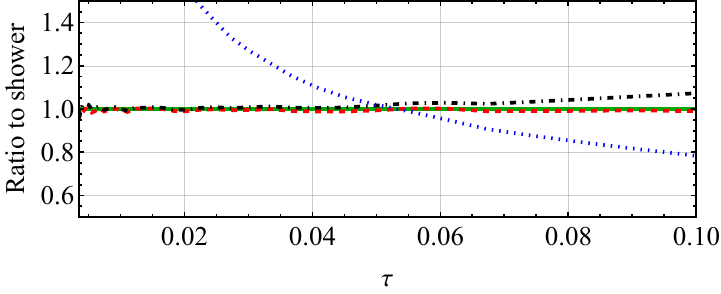}
\caption{\label{fig:shower-conj-cmp} Predictions for the thrust distributions at LL: shower (solid green), full $N$-space formula \eqref{eq:Sig_full_LL}
(dashed red), full $N$-space formula approximated with the saddle-point method in Eq.~\eqref{eq:Sig_full_SP_LL} (dotted black), and truncated $N$-space formula in Eq.~\eqref{eq:dSig_trunc_LL}
(dot-dashed blue). Ratios to the shower result are shown in the lower inset.}
\end{figure}

Results for the LL differential thrust spectrum are shown in Fig.~\ref{fig:shower-conj-cmp}. This numerical study confirms the exact
equivalence between the predictions obtained using the shower formulation (solid
green) and the full conjugate-space expression of Eq.~\eqref{eq:Sig_full_LL} (dashed red). The inverse Laplace transform involved in Eq.~\eqref{eq:Sig_full_LL} is
numerically challenging as the $N$-integrand, which is already a numerical
integral over $\bar\tau$ at fixed $N$, becomes highly oscillatory at large imaginary
values of $N$.
Therefore, as an additional check, we also employ the saddle-point approximation
\begin{equation}
\label{eq:Sig_full_SP_LL}
\frac{\rd }{\rd \tau}
\RresfullSPLL(\tau)
\, = \,
\frac{1}{2\pi} \,
e^{\widetilde{N}_{\tau} \tau-
\widetilde{R}_{\rm LL}(\widetilde{N}_{\tau}; \,\tau_{\rm L})}
\,
\sqrt{
\frac{2\pi}{\frac{\rd^{2}}{\rd N^{2}}
\big[\!
-\widetilde{R}_{\rm LL}
(\widetilde{N}_{\tau}; \, \tau_{\rm L})
\big]}}
\, ,
\end{equation}
that has been obtained by truncating the exponent at the second order around the
saddle-point $\widetilde{N}_{\tau}$. The latter is determined numerically by solving
the equation
\begin{equation}
\tau - \frac{\rd}{\rd N}
\widetilde{R}_{\rm LL}(\widetilde{N}_{\tau}; \,\tau_{\rm L})
\, = \, 0
\, ,
\end{equation}
at a given $\tau$ value.

The corresponding result is shown as a dotted
black line in Fig.~\ref{fig:shower-conj-cmp}. We observe very good agreement of the saddle-point approximation with respect to the full
conjugate-space result (dashed red) at small values of $\tau$, while deviations increase
towards larger $\tau$. This behavior is expected, since the saddle-point
approximation is formally valid in the asymptotic large-$N$ limit.

By contrast, the prediction obtained by truncating the radiator by means of the theta-function approximation (dot-dashed blue) exhibits a markedly different behaviour, particularly in the peak region. To study this effect, we include higher-order contributions in the truncated radiator, according to the formal expansion \cite{Catani:2003zt}
\beq
\label{eq:exp-truncation}
\int_{0}^{1}
\frac{\rd \bar{\tau}}{\bar{\tau}} \,
\Rpll(\bar{\tau}) \,
(e^{-N \bar{\tau}}-1)
& = &
-\Gamma
\left(1-\frac{\partial}{\partial \ln N }\right)
\Rll(1/N)
+
\mathcal{O}\left(1/N\right)
\nnb\\
& = &
\frac{1}{\as \beta_{0}} g_{1}(1/N) +
\gamma_{E} \, \frac{\rd }{\rd \lambda} g_{1}(1/N)
\nnb\\
&&
\quad
+ \,
\frac{\as \beta_{0}}2
\big(\gamma_{E}^{2}+ \zeta(2)\big)
\frac{\rd^{2} }{\rd \lambda^{2}} g_{1}(1/N) +
\cdots \, ,
\eeq
where in the last equality we have explicitly displayed the LL, NLL, and NNLL
contributions (with ellipses denoting higher logarithmic and power suppressed
contributions). Note that the terms containing powers of $\gamma_E$ could be systematically accounted for (and resummed at all orders) using $1/\overline{N}$ as opposed to  $1/N$ in the argument of $g_i$, see Eq.~\eqref{eq:theta_approx}.

\begin{figure}
  \centering
  \includegraphics[width=0.48\textwidth]{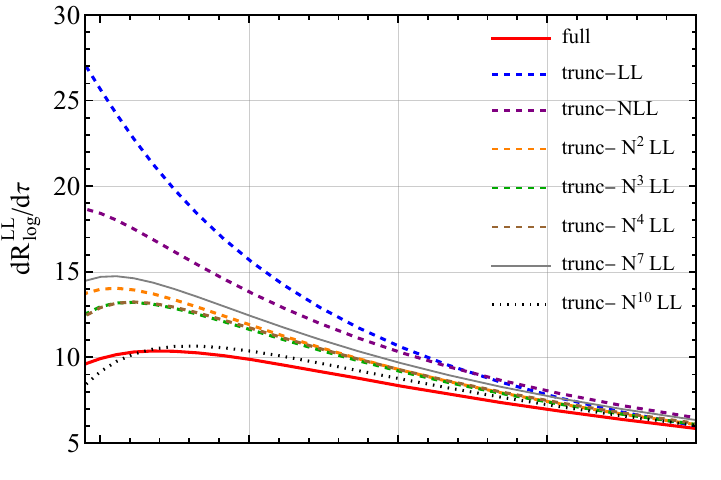}
  \hspace{0.5cm}
  \includegraphics[width=0.47\textwidth]{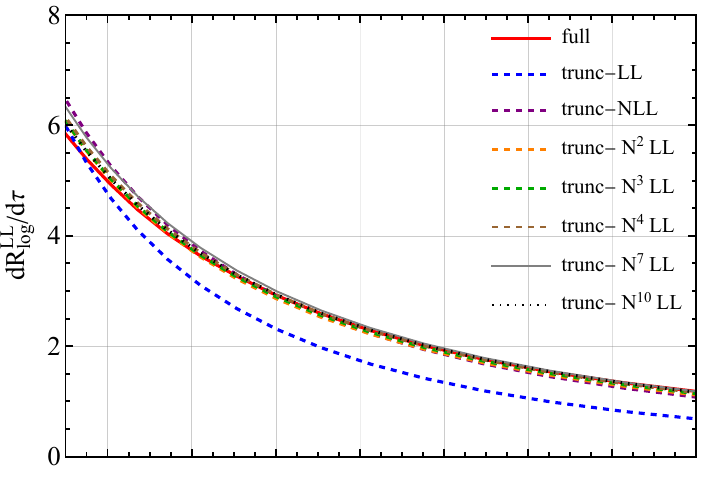}
  \\ \vspace{-0.2cm}
  \hspace{-0.08cm}\includegraphics[width=0.485\textwidth]{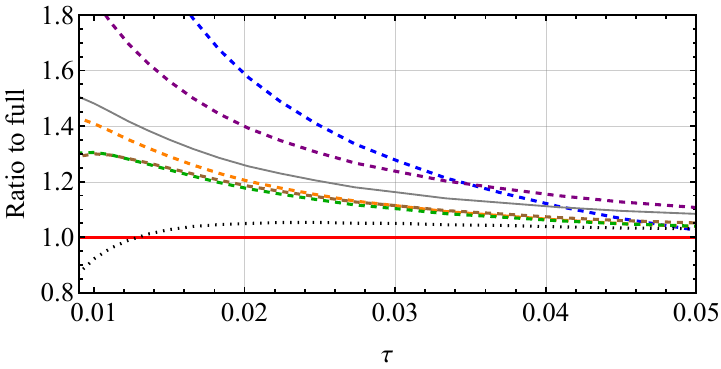}
  \hspace{0.1cm} \includegraphics[width=0.495\textwidth]{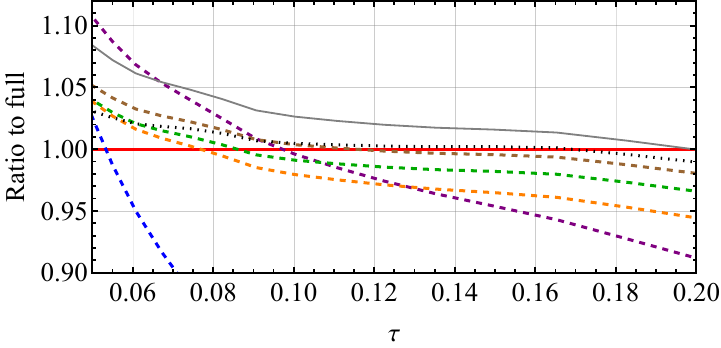}
  \caption{\label{fig:truncation-full-cmp} Comparison of the truncated $N$-space
    formula including increasingly higher-order N$^{k}$LL terms (dashed coloured
    curves up to N$^{4}$LL, gray solid for N$^{7}$LL and black dotted for  N$^{10}$LL) with the full result (red curve). Ratio to the full result is shown
    in the lower insets.}
\end{figure}

In Fig.~\ref{fig:truncation-full-cmp}, we illustrate the effect of including in
${\rd}\Rres^{\rm LL, trunc}(\tau)/{\rd \tau}$ terms beyond LL from the expansion
in Eq.~\eqref{eq:exp-truncation}. In the peak region (left panel), the inclusion
of higher-order terms up to N$^{4}$LL (dashed curves) shows a good convergence,
albeit to a curve that differs from the full result (red solid). If one includes
even higher-order terms, for example N$^{7}$LL (gray solid) and N$^{10}$LL
(black dotted), there are large variations, and the curves eventually approach
the full all-order result. However the convergence is rather slow, as each
successive term gives a sizeable contribution.

We stress that by inspecting the conjugate space N$^{k}$LL results with
$k=1,\dots,4$ shown in Fig.~\ref{fig:NKKN4LL} we would argue that the
convergence is good. On the other hand, the same terms shown in
Fig.~\ref{fig:truncation-full-cmp} up to N$^{4}$LL are contained also in the
results of Fig.~\ref{fig:NKKN4LL}. Thus, the convergence pattern in the latter
figure is not supported by the content of Fig.~\ref{fig:truncation-full-cmp}
when including higher orders.

The situation improves in the region at the right of the peak (right panel),
corresponding to larger values of $\tau$. Here, the truncated predictions
converge more rapidly towards the full result. Nevertheless, residual
oscillations at the level of a few percent persist also in this region, even
when very high-order terms are included.

Some comments are in order. The theta-function approximation was discussed in detail in the Beneke--Braun paper \cite{Beneke:1995pq} in the context of non-perturbative power
corrections to threshold resummation. Their analysis shows that
retaining the exponential leads to linear power corrections in
conjugate space for any value of $N$, that in turn can induce linear
power corrections to the total cross section once parton
luminosities are included.
By contrast, the theta-function approximation allows one to define the
radiator in conjugate space with non-perturbative effects confined to
the large-$N$ region. Adopting a suitable prescription for handling
the inverse Mellin transform (like the minimal
prescription \cite{Catani:1996dj,Catani:1996yz}) avoids the
introduction of any power-suppressed corrections. As linear power
corrections are argued to be absent from the total cross section,
in the context of threshold resummation it then seems more reasonable
to use the theta-function approximation as opposed to the full
exponential expression.

Conversely, in the case of thrust we know that linear power corrections are present in the final
result,
%Using the exponential expression implies linear corrections for all values of $N$ (with the exception of $N=0$, that corresponds to the total cross section).
thus we see no compelling physical justification for preferring the theta-function approximation over the exponential expression.
From the above numerical study we also note that subleading logarithmic corrections
to the theta-function approximation converge very slowly, especially at small values of  $\tau$.
This in turn implies that the choice of approach can lead to sizeable uncertainties, even when the expansion is truncated at relatively high orders.

\section{Numerical resummation in direct space}
\label{sec:num-resummation}
In this section we adopt a complementary perspective on the problem, and focus on numerical approaches to the resummation of multiple soft and/or collinear
emissions for the thrust distribution, formulated directly in momentum space.
For definiteness, we restrict ourselves to the first non-trivial logarithmic order, namely NLL accuracy.
The discussion builds upon the original CTTW resummation of
Ref.~\cite{Catani:1992ua},
based on the branching formalism, as well as on the automatic
numerical resummation frameworks CAESAR~\cite{Banfi:2003je,Banfi:2001bz},
ARES~\cite{Banfi:2014sua,Banfi:2018mcq} and
RadISH~\cite{Monni:2016ktx,Bizon:2017rah}.
These approaches can be seen as simplified parton showers with analytically controllable
logarithmic accuracy.

As for logarithmic accuracy, all shower-based formulations of thrust
resummation are formally equivalent to the approaches discussed previously,
namely resummation in conjugate space and its analytic inversion in direct space.
Differences may nevertheless arise at subleading level with respect to the
target accuracy. The purpose of this section is to investigate the similarities
and differences among these formulations, and to assess the numerical impact of
such subleading effects.

We begin by manipulating the resummation master formula Eq.~\eqref{eq:master-shower}, using the shorthand notation $\tau_i\equiv \tau(k_i)$.\footnote{Formula \eqref{eq:master-shower} is equal to Eq.~(2.6) of Ref.~\cite{Banfi:2014sua}, up to the finite virtual contributions $\mathcal{H}(Q^{2})$, which enter at NNLL accuracy.}
By imposing ordering in $\tau$ (i.e.~requiring that
$\tau_i\geq \tau_{i+1}$) and dropping the $1/m!$ factor, we get the equivalent
formula
\begin{equation}
\label{eq:master-shower-num0}
\Rres^{\rm NLL}(\tau)
\, = \,
\sum_{m=0}^{\infty}\int_T \,
\prod_{i=1}^{m} \,
[\rd k_{i}] \,
\big[M^{2}(k_{i})\big]_+
\Theta
\Big(\tau - \sum_{i=1}^{m}\tau_{i}\Big),
\end{equation}
where we have defined
\begin{equation}\label{eq:intTdef}
\int_T
\, \equiv \,
\int \prod_{i=1}^{m-1} \, \Theta(\tau_{i}-\tau_{i+1})
\, ,
\end{equation}
and $\tau_1$ denotes the hardest emission (i.e.~the one giving the largest contribution to $\tau$).
It is convenient to separate Eq.~(\ref{eq:master-shower-num0}) into the
sum of two terms, one with $\tau_1<\tau_0$ and one with $\tau_1>\tau_0$,
where $\tau_0$ is a small resolution cutoff. The first term does not depend upon $\tau$ (up to power corrections in $\tau_0/\tau$), and it is easily seen to exponentiate into
\begin{equation}
\label{eq:Runresolved}
[\Rres^{\rm NLL}]_{<\tau_0}
\, = \,
\exp
\left[
\int [\rd k] \,
\big[M^{2}(k)\big]_+
\Theta(\tau_0-\tau(k))
\right]
\, .
\end{equation}
Since the real contribution in the exponent is inhibited by the theta function,
the exponent is large and negative, owing to the virtual contribution, hence this term is strongly Sudakov suppressed. For simplicity, we do not show it in the following,\footnote{This term will actually appear again in the shower formulation discussed in the following, where it corresponds to the generation of no-emission events.} and focus on the contribution where there is at least one emission with $\tau>\tau_0$.

We then single out the $k_1$ integration, introduce a separation between
resolved and unresolved emissions, i.e.~emissions with $\tau_{i}>\epsilon\tau_{1}$ ($\tau_{i}<\epsilon\tau_{1}$),
with $\epsilon \ll 1$, and get
\begin{align}
\label{eq:master-shower-num1}
\Rres^{\rm NLL}(\tau)
\, = &
\int [\rd k_1] \, M^{2}(k_1)
\sum_{m=1}^{\infty}
\int_T \,
\bigg\{
\prod_{i=2}^{m} \,
[\rd k_{i}] \,
\big[M^{2}(k_{i})\big]_+ \,
\Big[
\Theta(\tau_{i} - \epsilon \tau_{1}) +
\Theta( \epsilon \tau_{1} - \tau_{i})
\Big]
\bigg\}
\notag \\
& \qquad
\times
\Theta
\Big(\tau - \sum_{i=1}^{m}\tau_i\Big), 
\end{align}
where it is understood that the integration over $k_1$ is limited to the region where $\tau_1>\tau_0$, and correspondingly $M^2(k_1)$ appears without `plus' distribution.
We remind the reader that the integration with a $T$ suffix is always as defined
in Eq.~\eqref{eq:intTdef}, i.e.~the factor $\Theta(\tau_1-\tau_2)$ is always included, even if $[\rd k_1]$ does not appear under the integral sign; moreover, an empty product  should be understood to be equal to 1, i.e. $\prod_{i=2}^{1}[\dots] \to 1$.

The combinatorics of Eq.~\eqref{eq:master-shower-num1}
is such that the full expression factorises
into the product of a term involving only $\tau_{i}<\epsilon\tau_{1}$ and another
involving only $\tau_{i}>\epsilon\tau_{1}$, leading to the final expression
\begin{align}
\label{eq:master-shower-num2}
\Rres^{\rm NLL}(\tau)
\, =  &
\int [\rd k_{1}] \,
M^{2}(k_{1}) \,
\exp
\bigg[
\int [\rd k] \,
\big[M^{2}(k)\big]_+
\Theta\big(\epsilon \tau_{1} - \tau(k)\big)
\bigg]
\notag \\
& \qquad
\times
\sum_{m=1}^{\infty}
\int_T \,
\bigg\{
\prod_{i=2}^{m} \,
[\rd k_{i}] \,
M^{2}(k_{i}) \,
\Theta(\tau_{i} - \epsilon \tau_{1})
\bigg\} \,
\Theta
\Big(\tau - \sum_{i=1}^{m}\tau_i\Big)
\, .
\end{align}
The factorisation and exponentiation
of the unresolved contribution can be understood by first observing that the sum in Eq.~\eqref{eq:master-shower-num1} would exponentiate
exactly if the measurement function $\Theta(\tau-\sum_i \tau_i)$ were omitted. Such an exponentiated expression would factorise into a product of two exponentials, associated with the unresolved and resolved contributions, respectively.
Furthermore, because of the $\tau$ ordering, the unresolved terms can be separated from the resolved ones. If $\epsilon$ is small enough they do not contribute to the measurement function (up to power corrections in $\epsilon\tau_1/\tau$), leading to a final expression where the unresolved contribution is exponentiated.

We can recast the single-radiation matrix element in terms of the Sudakov radiator introduced in Eq.~\eqref{eq:RNLL dir space},
\begin{equation}
\label{eq:Rdef}
\Rnll(\tau)
\, = \,
- \int [\rd k] \,
\big[M^{2}(k)\big]_+
\Theta( \tau - \tau(k))
\, = \,
\int [\rd k] \,
M^{2}(k) \,
\Theta( \tau(k) - \tau )
\, .
\end{equation}
Since $M^2(k)$ includes both real and
virtual contributions, as made explicit by the `plus' distribution, its unrestricted integral is finite due to the KLN cancellation. Neglecting this finite remainder, which is correct up to NLL order, justifies the second equality in Eq.~\eqref{eq:Rdef}.
Notice that $\Rnll(\tau)$ is a positive, monotonically decreasing function of $\tau$. Its logarithmic derivative $\Rpnll(\tau)$, defined as well in Eq.~\eqref{eq:RNLL dir space}, is also positive, and is single-logarithmic in $\tau$.
In terms of these quantities, we can finally recast Eq.~\eqref{eq:master-shower-num2} in the form
\begin{equation}
\label{eq:master-shower-num3}
\Rres^{\rm NLL}(\tau)
\, = \,
\int
\frac{\rd \tau_{1}}{\tau_{1}} \,
\Rpnll(\tau_{1}) \,
e^{-\Rnll(\tau_1)} \,
\mathcal{F}_{\tau_1;\tau}
\, = \,
\int
\rd \tau_1 \,
\frac{\rd e^{-\Rnll(\tau_1)}}{\rd \tau_{1}} \, \mathcal{\mathcal{F}}_{\tau_1;\tau}
\, ,
\end{equation}
where we have defined the transfer function
\begin{equation}
\label{eq:TransferDef}
\mathcal{F}_{\tau_1;\tau}
\, = \,
e^{\Rnll(\tau_1)-\Rnll(\epsilon \tau_{1})} \,
\sum_{m=1}^{\infty}
\int_T \,
\Big[
\prod_{i=2}^m \,
\frac{\rd \tau_{i}}{\tau_{i}} \,
\Rpnll(\tau_{i})
\Big] \,
\Theta(\tau_m-\epsilon \tau_1) \,
\Theta
\Big(\tau-\sum_{i=1}^m\tau_{i}\Big)
\, .
\end{equation}
Notice that, because of
$\tau$ ordering, the theta function of $\tau_m-\epsilon\tau_1$ implies that all the theta functions of $\tau_i-\epsilon\tau_1$ with $i<m$ are also satisfied, and thus can be omitted.
In Eq.~\eqref{eq:TransferDef} the logarithmic dependence on
the cutoff $\epsilon$ in the
integrals is compensated by the exponential prefactor.
We define the no-emission probability in the interval $[\tau',\,\tau'']$ as
\begin{equation}
\Delta_{\rm NLL}(\tau',\tau'')
\, \equiv \,
\exp\big[\Rnll(\tau')-\Rnll(\tau'')\big] \,
\Theta(\tau'-\tau'')
\, .
\end{equation}
Using the identity
\begin{align}
e^{\Rnll(\tau_1)-\Rnll(\epsilon \tau_{1})}
& = \,
\bigg[
\prod_{i=2}^m \,
e^{\Rnll(\tau_{i-1})-\Rnll(\tau_{i})}
\bigg]
\times
e^{\Rnll(\tau_m)-\Rnll(\epsilon\tau_1)} \notag \\
& = \,
\bigg[
\prod_{i=2}^m \,
\Delta_{\rm NLL}(\tau_{i-1},\tau_{i})
\bigg]
\times
\Delta_{\rm NLL}(\tau_m,\epsilon\tau_1)
\, ,
\end{align}
the transfer function can be rewritten as
\begin{equation}
\label{eq:master-transfer}
\mathcal{F}_{\tau_1;\tau}
\, = \,
\sum_{m=1}^{\infty}
\int
\Big[
\prod_{i=2}^{m} \,
\rd \tau_i \,
\frac{\rd \Delta_{\rm NLL}(\tau_{i-1},\tau_{i})}{\rd \tau_{i}}
\Big]
\times                        
\Delta_{\rm NLL}(\tau_{m},\epsilon \tau_{1})
\,
\Theta\Big(\tau-\sum_{i=1}^m\tau_{i}\Big)
\, .
\end{equation}
The above equation can be implemented as a shower procedure, defined by the formal
equations
\beq
\label{eq:master-transfer-algo}
\mathcal{F}_{\tau_1;\tau}
& = &
\mathcal{F}_{\tau_1;\tau}^{(1)}
\, ,
\\
\mathcal{F}_{\tau_1;\tau}^{(i)}
& =&
\Delta_{\rm NLL}(\tau_i,\epsilon\tau_1)
\times
\Theta\Big(\tau-\sum_{j=1}^i\tau_{j}\Big)
+
\int_{\epsilon \tau_1}^{\tau_i}
\rd \tau_{i+1} \,
\frac{\rd\Delta_{\rm NLL}(\tau_i,\tau_{i+1})}{\rd\tau_{i+1}} \,
\mathcal{F}^{(i+1)}_{\tau_1,\tau}
\, ,
\nnb
\eeq
that by iteration reproduces Eq.~\eqref{eq:master-transfer}. We immediately recognise that,
were it not for the $\Theta$ function, each $\mathcal{F}_{\tau_1;\tau}^{(i)}$ would be the sum of the probability of having no further emissions after the $i^{\rm th}$ one (first term) plus the probability to have subsequent emissions (second term), thus it would be one.
We also note that, in the leading-logarithmic approximation, each secondary emission with $i>1$ satisfies $\tau_{i}\ll \tau_1$.
As a result, the $\Theta$ function fully factorises from $\mathcal{F}_{\tau_1;\tau}$, and the latter just reduces to $\Theta(\tau-\tau_1)$. It follows that
the full expression of $\mathcal{F}_{\tau_1;\tau}$ becomes necessary only at the NLL
level, and thus can be evaluated in the LL approximation.
Thus the starting point of our analysis is represented by Eqs.~\eqref{eq:master-shower}, \eqref{eq:master-transfer}, and \eqref{eq:master-transfer-algo}, where we consistently replace $\Delta_{\rm NLL}$ with its LL counterpart
$\Delta_{\rm LL}(\tau_{i-1},\tau_{i}) \, = \, e^{\Rll(\tau_{i-1})-\Rll(\tau_{i})}$.

Algorithmically, Eq.~\eqref{eq:master-transfer-algo} can be implemented as a simplified parton
shower ordered in $\tau$. The hardest emission $\tau_{1}$
is generated by solving
\begin{equation}
\Delta_{\rm NLL}(1,\tau_1)
\, \equiv \,
e^{-\Rnll(\tau_{1})}
\, = \,
r_{1}
\, ,
\end{equation}
while subsequent emissions are obtained from
\begin{equation}
\label{eq:ShowerStep}
\Delta_{\rm LL}(\tau_{i-1},\tau_{i})
\, = \,
r_{i}
\, ,
\end{equation}
with $r_{i}\in(0,1)$ uniformly distributed random numbers. If $\tau_{i}<\epsilon\tau_{1}$ the
iteration is stopped.
The resulting emissions are then combined to give the observable $\tau$
according to $\tau=\sum_{i}\tau_{i}$.

At NLL accuracy one may further assume $\tau_{i}\sim\tau_{1}$ for $i>1$.
In fact, for the first $i>1$ such that $\tau_i\ll \tau_1$, the transfer function
$\mathcal{F}_{\tau_1;\tau}^{(i)}$ reduces to $\Theta(\tau-\sum_{j=1}^{i-1}\tau_j)$ (see Eq.~\eqref{eq:master-transfer-algo}) and thus only terms with $\tau_i \sim \tau_1$
survive.
Therefore, we can approximate
\begin{equation}
\Delta_{\rm LL}(\tau_{i-1},\tau_{i})
\, = \,
e^{-\Rpll(\tau_{1})\ln\frac{\tau_{i-1}}{\tau_{i}}}
\, + \,
\mathcal{O}({\rm NNLL})
\, ,
\end{equation}
that in the shower context leads to the analytic solution
of Eq.~\eqref{eq:ShowerStep}
\begin{equation}
\tau_{i} = \tau_{i-1} \exp\!\left(\frac{\ln r_{i}}{\Rpll(\tau_{1})}\right)\,.
\end{equation}
The transfer function (see Eq.~\eqref{eq:TransferDef}) reduces to
\begin{align}
\label{eq:radish transfer function}
\mathcal{F}_{\tau_1;\tau}
& = \,
e^{-\Rpll(\tau_{1})\ln\frac{1}{\epsilon}}
\sum_{m=1}^{\infty} \,
\prod_{i=2}^{m}
\left[
\, \int
\frac{\rd \tau_{i}}{\tau_{i}} \,
\Rpll(\tau_{1}) \,
\Theta(\tau_{i-1}-\tau_i)
\right]
\Theta(\tau_m-\epsilon\tau_1) \,
\Theta\Big(\tau-\sum_{i=1}^{m}\tau_{i}\Big)
\notag \\
& = \,
e^{-\Rpll(\tau_{1})\ln\frac{1}{\epsilon}}
\sum_{m=1}^{\infty}
\frac{\big[\Rpll(\tau_{1})\big]^{m-1}}{(m-1)!}
\prod_{i=2}^{m} \int_{\epsilon \tau_{1}}^{\tau_{1}}
\frac{\rd \tau_{i}}{\tau_{i}} \,
\Theta\Big(\tau-\sum_{i=1}^{m}\tau_{i}\Big)
\, ,
\end{align}
which corresponds to the RadISH formulation of thrust resummation.

Finally, for thrust one may further exploit the fact that $\tau_{1}\sim\tau$ at NLL accuracy,
which is motivated as before.
In this case we can expand the Sudakov exponents in the transfer function
around $\tau$, as opposed to $\tau_1$,
and obtain (see Eq.~\eqref{eq:master-shower-num3})
\begin{equation}
\label{eq:finalthrustcaesar0}
\Rres^{\rm NLL}(\tau)
\, = \,
\int^\tau
\!\!
\rd\tau_1 \,
\frac{\rd e^{-\Rnll(\tau_1)}}{\rd \tau_1} \,
e^{-\Rpll(\tau)\ln\frac{1}{\epsilon}} \,
\sum_{m=1}^{\infty}
\frac{\big[\Rpll(\tau)\big]^{m-1}}{(m-1)!}
\prod_{i=2}^{m}
\int_{\epsilon\tau_1 }^{\tau_1}
\frac{\rd \tau_{i}}{\tau_{i}} \,
\Theta\Big(\tau-\sum_{i=1}^{m}\tau_{i}\Big)
\, .
\end{equation}
Using the approximation
\begin{equation}
\Rnll(\tau_1)
\, =\,
\Rnll(\tau) \, + \,
\Rpll(\tau) \, \log\frac{\tau}{\tau_1} \, + \,
{\cal O}({\rm NNLL})
\, ,
\end{equation}
and rescaling $\tau_{1}=\rho\,\tau$,
$\tau_{i}=\rho_{i}\,\rho\,\tau$, one can easily carry out analytically the integration over $\rho$, and
the cumulative distribution becomes
\begin{equation}
\label{eq:finalthrustcaesar}
\Rres^{\rm NLL}(\tau)
\, = \,
e^{-\Rnll(\tau)} \,
e^{-\Rpll(\tau)\ln\frac{1}{\epsilon}} \,
\sum_{m=1}^{\infty}
\frac{\big[\Rpll(\tau)\big]^{m-1}}{(m-1)!}
\bigg[
\prod_{i=2}^{m} \,
\int_{\epsilon}^{1}
\frac{\rd \rho_{i}}{\rho_{i}}
\bigg] \,
e^{-\Rpll(\tau)\ln(1+\sum_{i=2}^{m}\rho_{i})}
\, .
\end{equation}
When carrying out the $\rho$ integration, we should remember that formally it has a lower bound $\rho> \tau_0/\tau$, and that the ensuing $\tau_0$ dependence should be compensated
by the term in Eq.~\eqref{eq:Runresolved}. On the other hand, having performed our expansions, we see that the lower integration bound can be taken to zero.

In the case of thrust one can carry out all of the remaining integrations in Eq.~\eqref{eq:finalthrustcaesar} analytically, by means of an integral transform following the steps around Eqs.~(3.23)-(3.26) of Ref.~\cite{Banfi:2004yd}. This yields the CAESAR/ARES result at NLL:
\begin{equation}
\Rres^{\rm NLL}(\tau)
\, = \,
e^{-\Rnll(\tau)} \,
\frac{ e^{-\gamma_{E} \Rpll(\tau)}}{\Gamma\big(1+\Rpll(\tau)\big)}
\, .
\end{equation}
This result coincides with the analytic inversion of the Laplace-space
resummation at NLL accuracy.  Let us observe that this equivalence
holds in spite of the fact that in the present case we did not use the
theta-function approximation which is instead used in the analytic
direct-space formulation.

\begin{figure}[htb]
\centering
\includegraphics[width=0.6\textwidth]{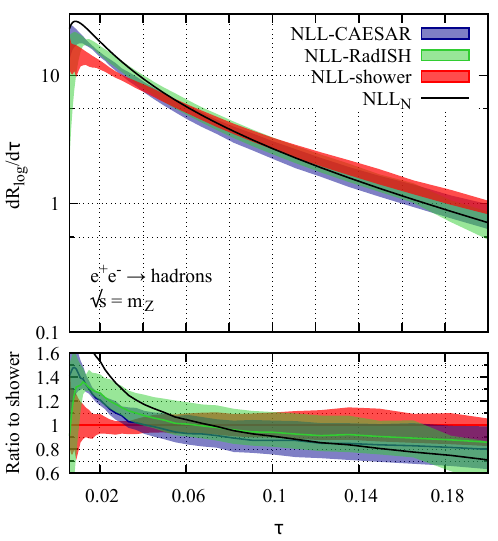}
\caption{Comparison of NLL results obtained with CAESAR/ARES (blue), RadISH (green), shower (red), and the NLL variant of the conjugate-space resummation
  formula~(\ref{eq:directdiff}) (black).
\label{fig:nll-cmp}}
\end{figure}

In Fig.~\ref{fig:nll-cmp}, we compare NLL resummed predictions for the thrust
distribution obtained using the three formulations discussed above, which are
equivalent according to the logarithmic counting. By ``shower'' (red) we
indicate the use of Eq.~\eqref{eq:master-transfer-algo}, ``RadISH'' (green)
denotes Eq.~\eqref{eq:radish transfer function}, while ``CAESAR'' (blue)
corresponds to Eq.~\eqref{eq:finalthrustcaesar}. The associated uncertainty
bands are estimated from uncorrelated variations of the resummation scale $\muL$
and the renormalisation scale $\muR$ by a factor of two above and below their
central values, always keeping $1/4 < \muL/\muR < 4$. For comparison, we also
show the corresponding conjugate-space NLL result in black, which, we stress
again, includes the theta-function approximation. For $\tau \gtrsim 0.03$, the
RadISH central prediction lies about $5\%$ above the corresponding CAESAR
result, although their uncertainty bands largely overlap. This indicates that
the expansions around $\tau_{1}$ (RadISH) and around $\tau$ (CAESAR) are
formally equivalent up to subleading corrections. The RadISH prediction exhibits
a slightly better agreement with conjugate-space resummation in the $\tau$
window $[0.02, 0.06]$. The shower prediction instead displays a harder spectrum.
At large values of $\tau$, the RadISH prediction lies at the lower edge of the
shower uncertainty band, whose width amounts to roughly $15$--$20\%$.

\begin{figure}
  \centering
  \hspace{-0.1cm}\includegraphics[width=0.48\textwidth]{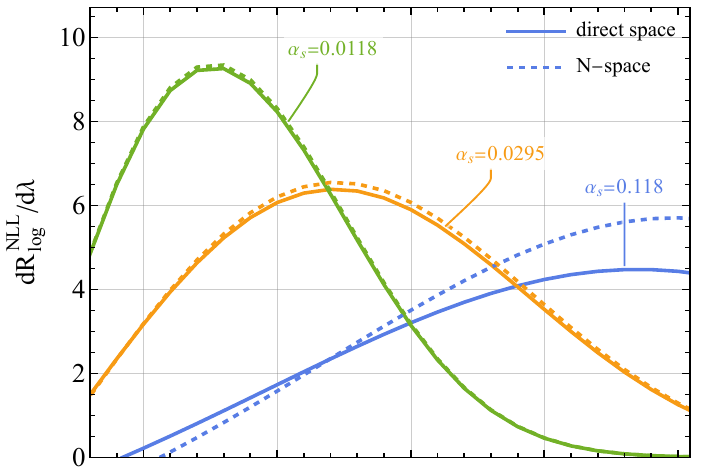}
  \hspace{0.3cm}
  \includegraphics[width=0.48\textwidth]{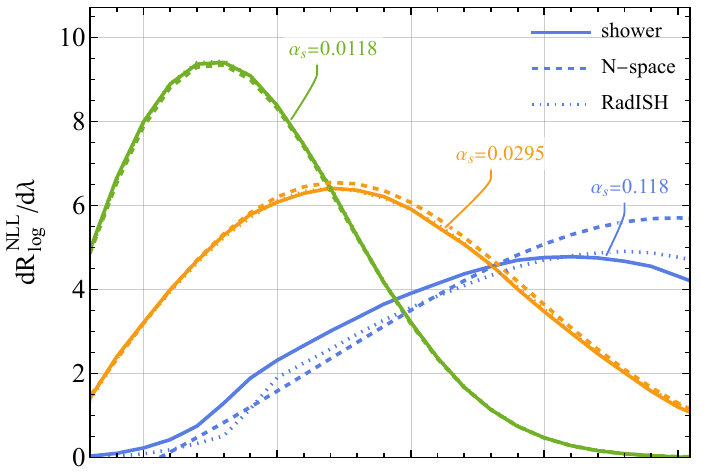}
  \\ \vspace{0cm}
  \hspace{-0.1cm}\includegraphics[width=0.48\textwidth]{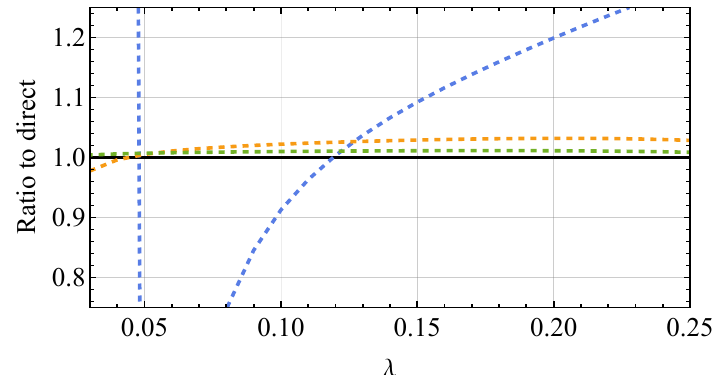}\hspace{0.4cm}
  \includegraphics[width=0.48\textwidth]{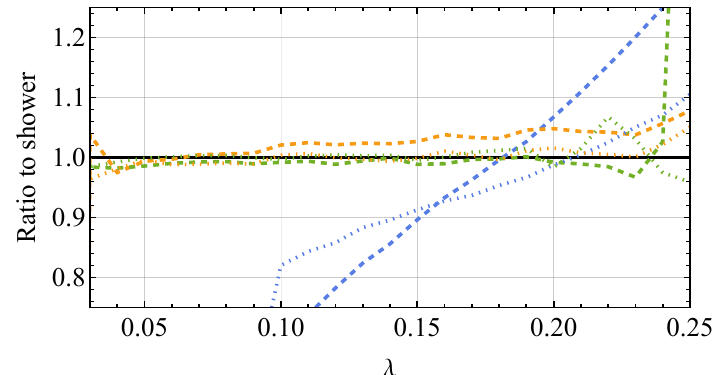}
  \caption{\label{fig:small_as} Comparison of different NLL predictions in the small-$a_s$ limit as a function of $\lambda_\tau$.  Left panel: direct- and
    conjugate-space resummations. Right panel: shower, RadiSH and conjugate-space results. Ratio plots are shown in the bottom panels.}
\end{figure}

As an additional numerical check that the observed differences among the
various approaches are logarithmically subleading, we consider the
$a_s \to 0$ limit at fixed $\lambda_\tau = a_s \,\beta_{0} \,\log 1/\tau$. To this end, we
compute the spectrum as a function of $\lambda_\tau$ while progressively decreasing the $a_s$ value. The results are shown in Fig.~\ref{fig:small_as}. As
expected, all formalisms converge to the same result as $a_s$ becomes
increasingly small.

It would be interesting to extend this analysis to higher logarithmic accuracy. We postpone this to future work; in any case, we do not expect our conclusions to be significantly affected.

\section{Conclusions}\label{sec:conclusions}

\textbf{Motivations.}
In this work, we have investigated the impact of certain subleading terms of
kinematic origin that arise in the logarithmic resummation of the thrust shape
variable in the $e^+e^-\to\,$hadrons process. One motivation for this study was
the finding of Ref.~\cite{Aglietti:2025jdj}, where substantial differences were
noted when computing the thrust distribution in the so-called ``conjugate''
space, as compared to the computation performed in thrust (or direct) space.
Conjugate space is defined such that the factorisation of the kinematics for the
emission of soft-collinear gluons is exactly preserved, whereas in direct space
this factorisation holds only up to the logarithmic accuracy at which one is
working. A second motivation came from Ref.~\cite{Catani:1996yz}: in the
resummation of threshold-enhanced corrections in hadronic collisions, it was
found that the direct-space approach leads to the generation of terms that,
although subleading in the logarithmic sense, display a factorial growth with
the perturbative order, and thus give rise to spurious power-suppressed
ambiguities\footnote{In the case of gluon-initiated processes, these
  ambiguities are of order $(\Lambda_{\rm QCD}/Q)^{(0.16)}$~\cite{Catani:1996yz}, which is barely
  distinguishable from a constant.}.

\textbf{Content of the study.} In this work we have firstly investigated the
ambiguities that arise when transforming the resummation formulae from conjugate
to direct space, and vice versa. Secondly, we have highlighted, for the
first time, another potential source of ambiguity, which arises in the
derivation of conjugate-space resummation from the replacement
\begin{equation}
\label{eq:app-thetaApprox}
1-e^{-N\tau}
\, = \,
\Theta
\Big(
\tau - \frac{1}{Ne^{\gamma_{E}}}
\Big)
\, + \,
{\cal O}({\rm NNLL})
\, .
\end{equation}
Such a replacement was proposed since the seminal papers on the resummation of
shape variables \cite{Catani:1992ua}. It was further recommended in
Ref.~\cite{Beneke:1995pq} as a mean to avoid the presence of
${\cal O}(\Lambda_{\rm QCD}/Q)$ corrections (where $\Lambda_{\rm QCD}$ is a
typical hadronic scale, and $Q$ is the scale of the process) when performing
threshold resummation for the Drell-Yan process, in view of the arguments that
disfavour the presence of such corrections in the Drell-Yan total cross section.
On the contrary, it is in general believed that linear
power corrections are present in the thrust distribution, so that no strong
motivation remains for the approximation of Eq.~(\ref{eq:app-thetaApprox}).

Finally, we have examined the relation between conjugate- and direct-space
formalisms and numerical approaches to the resummation of event-shape variables,
such as those of
Refs.~\cite{Banfi:2003je,Banfi:2001bz,Banfi:2014sua,Banfi:2018mcq,Monni:2016ktx,Bizon:2017rah}.

\textbf{Results: Laplace inversion.}
% We have analysed the towers of subleading
% terms arising from the inversion of the conjugate-space resummation formula at a
% fixed nominal logarithm accuracy.
We have started with the resummation formula in conjugate space, truncated at a
given logarithmic accuracy, and performed its inverse Laplace transform to
direct space. This generates an infinite tower of subleading corrections. In
case our nominal conjugate-space accuracy is at the double-logarithmic (DL) level,
the tower of subleading corrections is a series with an infinite radius of
convergence, and we do not expect any problems by truncating it. In fact, we
have demonstrated that by including the first subleading corrections we achieve
a very good approximation to the resummation of the whole series of subleading
terms.

Moving on to the leading-logarithmic (LL) approximation, we have found that the
expansion of subleading corrections has zero radius of convergence, i.e.~it is
to be interpreted as an asymptotic expansion. The growth of the perturbative
coefficients is mild, proportional to the logarithm of the order, which we
denote as ``log-factorial'' growth. This behaviour is to be contrasted with
other asymptotic expansions where the growth of the coefficients is proportional
to the order itself, which leads to factorial coefficients.

We estimate that the ambiguity related to the truncation of the log-factorial
series is suppressed as an exponential of $-\tau Q/\Lambda_{\rm QCD}$, as opposed to the
power-like ambiguity $(\Lambda_{\rm QCD}/Q)^a$ of factorially-growing expansions.
Nevertheless, the convergence of the series is relatively slow. The asymptotic
nature of the series is physically due to the presence of the Landau pole in the
LL resummation formula, which is absent in the double-logarithmic approximation.
This slow convergence persists when we consider higher-order corrections to the
resummed result, namely the N$^k$LL approximation with $k$ up to four.

We have also considered the opposite transformation, from a truncated DL
resummation formula in direct space to conjugate space. We find that the series
of subleading logarithmic terms generated in conjugate space is not convergent,
with coefficients that grow factorially. In other words, if we decided to fit
the Laplace moments of the observed distribution, we should be prepared to
account for much larger theoretical ambiguities of kinematic origin. This fact
highlights that the absence of full factorisation of soft-collinear emissions in
direct space remains a problem that is bound to show up under appropriate
circumstances.

% Overall, we find that the conjugate-space result has slightly better convergence
% properties than the direct-space one.
Direct- and conjugate-space results display sizeable differences for
$\tau \lesssim 0.02$, i.e.~at the peak of the distribution. Above the peak
region, the two predictions at the highest available logarithmic accuracy, N$^{4}$LL, differ by few percent, both in magnitude and in their slopes. After
applying a matching procedure to correct the resummation for fixed-order effects
up to N$^3$LO, we find that the direct- and the conjugate-space results differ
at the level of 5\% in the region usually considered for fits of the strong
coupling, i.e.~$\tau \in [0.06,0.15]$. This difference is covered by the
theoretical error estimate only with suitably conservative scale variations.
However, it is not as strong as the one reported in
Ref.~\cite{Aglietti:2025jdj}, where it exceeds 10\%.

\textbf{Results: theta-function approximation.} We have examined the impact of
the theta-function approximation (\ref{eq:app-thetaApprox}) that is usually
adopted when deriving the conjugate-space resummation formula. To this aim, we
have first considered a basic resummation formula, where the multiple-radiation
matrix element is factorised as a product of single-emission terms, and the
thrust variable $\tau$ is as well expressed as a sum of individual
contributions. We have compared this formula with the resummation formula that
employs Eq.~\eqref{eq:app-thetaApprox}, which in turn, by means of a Laplace
transform, leads to the conjugate-space formulation.
Our analysis, performed at the LL level, shows that using
Eq.~\eqref{eq:app-thetaApprox} has a significant impact on the shape of the
thrust distribution, yielding a considerable softening and a strong enhancement
of the peak region. This effect acts in the opposite direction to the
transformation from conjugate to direct space, which instead lowers the
distribution in the peak region and produces an overall hardening.

As a further point, we have noticed that by adding subleading terms to
formula~(\ref{eq:app-thetaApprox}) leads to an expansion that displays a very
slow convergence. In particular, results that include corrections to the
theta-function approximation up to N$^{4}$LL seem to progressively stabilise,
although not around the true asymptotics. The latter is reached
only upon including extremely high orders, in our case N$^{10}$LL. The same
corrections to the theta-function approximation enter the standard
conjugate-space predictions order by order in the logarithmic counting.
Therefore, this analysis casts some doubts on the apparent convergence that is
observed in the conjugate-space results up to N$^{4}$LL around the thrust peak.

\textbf{Results: numerical resummations.} Finally, we have considered numerical
resummation approaches such as CEASAR/ARES, RadISH, and a parton-shower
algorithm, limiting ourselves to the NLL accuracy level. We find that the
CAESAR/ARES and the RadISH predictions are compatible with the conjugate-space
result for $\tau \gtrsim 0.03$, namely in the region above the peak of the
distribution. We have shown that the shower approach preserves exactly the
kinematics of the emissions and yields results equivalent to those obtained in a
conjugate space approach in which the theta-function approximation is not used.
It yields a rather different shape with respect to the standard conjugate-space
result, with a suppressed peak region and a harder tail. All methods, as
expected, become more compatible with each other for smaller values of $\as$.
Resummation performed using the SCET framework is yet another alternative
relatively similar to the direct space approach. The two methods can be related,
see e.g. Ref.~\cite{Almeida:2014uva}. We therefore believe that the differences
identified in this study among the evaluated methods reflect inherent
ambiguities present across the full range of possible approaches.

\textbf{Concluding remarks.}
In this study we have found that different legitimate choices in the resummation methods
often lead to differences in the result that are not covered by standard scale
variations. This fact suggests that, when fitting the thrust distribution using
resummation-improved calculations, more conservative estimates of the
theoretical errors should be adopted.

\section*{Acknowledgements}
We thank G. Ferrera and P.F. Monni for useful discussions. We
are grateful to U. Aglietti, G. Ferrera, J. Miao, P.F. Monni, G. Salam and G.
Zanderighi for comments on the manuscript. L.R. thanks M. Bonvini for useful
discussions on the Borel prescription. The work of L.B. is funded by the
European Union (ERC, grant agreement No. 101044599, JANUS). Views and opinions
expressed are however those of the authors only and do not necessarily reflect
those of the European Union or the European Research Council Executive Agency.
Neither the European Union nor the granting authority can be held responsible
for them. P.T. has been partially supported by the Italian Ministry of
University and Research (MUR) through grant PRIN\_2022BCXSW9.

\appendix
\section{Resummation ambiguities}\label{sec:resummationAmbiguities}

The fixed-order truncation of an asymptotic series in the strong coupling
inevitably leads to the appearance of an ambiguity, ultimately connected
with the scale $\Lambda_{\rm QCD}$ at which the strong coupling becomes
non-perturbative.
We want to estimate how this ambiguity scales with the relevant hard scale
$Q_{h} = \tau Q$, with $Q$ the centre-of-mass energy of the
electron--positron collision. We denote by $q_n$ the $n$-th term of
the series. The minimum term $q_{\bar{n}}$ is defined by the condition
$q_{\bar{n}} = q_{\bar{n}-1}$, i.e.~the value of $n$ where two subsequent
terms become equal.
Truncating the series at $q_{\bar{n}}$ induces an uncertainty of the order
of the leftover term, from which the scaling of the ambiguity with $\tau Q$
can be inferred.

In our case, we consider the series of coefficients
\beq
q_n
\, = \,
b^n \, {\rm Lf}(n)
\, ,
\qquad
b
\, \equiv \,
\frac{\as(Q)}{1-2\lambda_\tau}
\, ,
\eeq
which corresponds to the $D_n$ coefficient in Eq.~\eqref{eq:LLasymptotics} with the
modulation factor $P_n$ removed.
Using the definition $\lambda_\tau=a_s(Q)\beta_0 \log(1/\tau)$,
and the
lowest-order running coupling $a_s(Q)=1/(\beta_0 \log(Q^2/\Lambda_{\rm QCD}^2))$,
one can easily show that
\beq
b
\, = \,
\as(\tau Q)
\, .
\eeq
The minimum of $q_n$ is found by imposing $q_{\bar{n}} =
q_{\bar{n}-1}$, which leads to
\beq
\label{eq:min_series}
b \, \log \bar{n} \, = \, 1
\, .
\eeq
We define the resummation ambiguity $\delta(\tau Q;\Lambda_{\rm QCD})$ as
\begin{equation}
\delta(\tau Q;\Lambda_{\rm QCD})
\, = \,
q_{\bar{n}}
\, = \,
(b \log \bar{n})^{\bar{n}}
\times
\frac{{\rm Lf}(\bar{n})}{\log^{\bar{n}} \bar{n}}
\, = \,
\frac{{\rm Lf}(\bar{n})}{\log^{\bar{n}} \bar{n}}
\, .
\end{equation}
To analyse the energy scaling of $\delta(\tau Q;\Lambda_{\rm QCD})$, we study
the large-$n$ behaviour of the log-factorial function.
Its asymptotics can be derived along the same lines as Stirling’s approximation
for the standard factorial:
\beq
\log [{\rm Lf}(n)]
& = &
\log\Big(\prod_{k=2}^n \log k\Big)
\, = \,
\sum_{k=2}^n \log(\log k)
\sim
\int_1^{n+\frac12} {\rm d}x \, \log(\log x)
\nnb\\
& = &
\Big(n+\frac12\Big)
\log\!\Big(\log\!\big(n+\tfrac12\big)\Big)
- {\rm Li}\!\left(n+\tfrac12\right)
+ \gamma_E
\, ,
\eeq
where ${\rm Li}(t) = \int_0^t {\rm d}x/\log x$ is the logarithmic integral,
with the singularity at $x=1$ regulated in the principal-value sense.
Expanding at large $n$ gives
\beq
\log[{\rm Lf}(n)]
\, \sim \,
n \log(\log n)
-
\frac{n}{\log n}
+ \dots
\, .
\eeq
Hence,
\beq
\delta(\tau Q;\Lambda_{\rm QCD})
\, = \,
\frac{{\rm Lf}(\bar{n})}{\log^{\bar{n}} \bar{n}}
\, \sim \,
\exp\left(-\frac{\bar{n}}{\log \bar{n}}\right)
\, .
\eeq
The minimum condition \eqref{eq:min_series} implies
\beq
\bar{n}
\, = \,
e^{1/b}
\, = \,
e^{1/\as(\tau Q,\Lambda_{\rm QCD})}
\, = \,
\left(\frac{\tau Q}{\Lambda_{\rm QCD}}\right)^{2\beta_0/\pi}
\, .
\eeq
Thus the ambiguity becomes
\begin{equation}
\delta(\tau Q;\Lambda_{\rm QCD})
\, \sim \,
\exp\!
\left[
-\,
\frac{(\tau Q/\Lambda_{\rm QCD})^{\frac{2\beta_0}{\pi}}}
{{\frac{2\beta_0}{\pi}}\log(\tau Q/\Lambda_{\rm QCD})}
\right]
\, .
\end{equation}
This expression is exponentially suppressed at large $\tau Q$, i.e.~the
resummation ambiguity decreases faster than any power of
$\Lambda_{\rm QCD}/(\tau Q)$.
This behaviour stems directly from the mild (logarithmic) growth of the
coefficients $q_n$, and can be contrasted with the power-like
$(\Lambda_{\rm QCD}/\tau Q)^a$ scaling that follows from a plain factorial
growth \cite{Catani:1996yz}.

\begin{figure}
\centering
\includegraphics[width=0.5\textwidth]{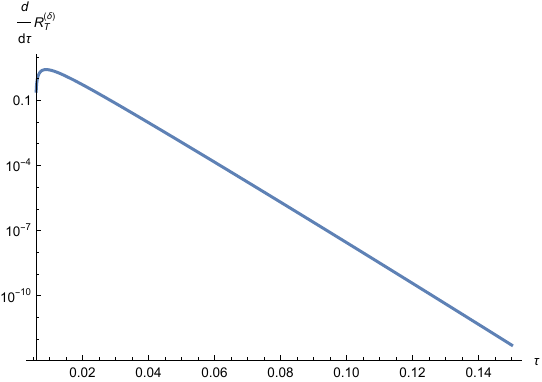}
\caption{\label{fig:res-ambiguity}
Log-plot of ${\rm d} {\cal R}^{(\delta)}_T / {\rm d}\tau$ as a function of $\tau$,
for fixed $Q = 91\,$GeV, and with $\Lambda_{\rm QCD} = 0.2$\,GeV.}
\end{figure}

In Fig.~\ref{fig:res-ambiguity}, we show the logarithmic plot of
\beq
\frac{{\rm d}}{{\rm d}\tau} {\cal R}^{(\delta)}_T(\tau)
\, \equiv \,
\frac{{\rm d}}{{\rm d}\tau}
\left[
e^{\frac{1}{a_s \beta_0} g_1(\lambda_{\tau})}
\,
\delta(\tau Q;\Lambda_{\rm QCD})
\right]
\, ,
\eeq
which quantifies the impact of the ambiguity on the $\tau$ distribution at fixed
$Q$.
We observe that the effect reaches its maximum around $\tau \sim 0.01$, close
to the peak of the distribution, and becomes exponentially suppressed for larger
values of~$\tau$.

We stress that to obtain Fig.~\ref{fig:res-ambiguity} several potentially
important factors were neglected, thus the figure provides only an
indication of the shape of the ambiguity. In particular, one should
include a factor accounting for the ambiguity in the precise location
of $\bar{n}$, which becomes particularly important in the case of an
asymptotic divergence of constant sign (see, for instance, footnote 1
in Ref.~\cite{Beneke:2016cbu}). In the present case, since we have an
oscillating overall factor, this ambiguity should correspond to a
constant factor of the order of the period of the oscillating phase.
\section{Borel Prescription}
\label{app:borel}
The Borel prescription \cite{Forte:2006mi,Abbate:2007qv} provides an alternative to the minimal prescription \cite{Catani:1996dj,Catani:1996yz} for numerically performing the Laplace inversion while avoiding the Landau-pole singularity.
In the context of hadronic collisions, the difference between the two prescriptions  has been found to be in most cases negligible, unless the value of $\tau^{\rm had} = Q^2/s$ is close to the Landau pole of the strong coupling $\tau^{\rm had} = 1 - \Lambda_{\rm QCD}^2/Q^2$ \cite{Bonvini:2010tp}.
In the case of thrust resummation, however, the sensitivity to the Landau pole is relevant for values of $\tau$ around the peak of the distribution, which motivates studying the impact of the chosen Landau-pole prescription on the results.

The Borel prescription was originally formulated in Refs.~\cite{Bonvini:2010tp,Bonvini:2012sh} in the context of threshold resummation. It
can be adapted to the thrust case
by noticing that the
%Laplace transform,
%\begin{align}
%\widetilde f(N)
%\, = \,
%\int_0^\infty \rd \tau \,
%e^{-N\tau} \, f(\tau)
%\, ,
%\end{align}
%and the
Mellin transform,
\begin{align}
\mathcal M[f] (N)
\, = \,
\int_0^1 \rd x \,
x^{N-1} \, f(x)
\, ,
\end{align}
can be recast as a Laplace transform in variable $\tau$ by means of the change of variable $x = e^{-\tau}$ as long as $f(x)$ is defined in the range $ 0 < x < 1$.
%As a consequence, we can immediately use the formulae deduced for threshold resummation in Ref.~\cite{Bonvini:2012sh}.

We first identify the differential $\tau$ spectrum as the function to be Laplace transformed,
\begin{align}
\frac{\rd \Rres  (\tau)}{\rd \tau} 
\, = \,
\frac{1}{2 \pi i}
\int_{c - i \infty}^{c+ i \infty}
\rd N \,
e^{N\tau} \,
\widetilde {\mathcal{R}}_{\rm \scriptscriptstyle log} (\lambda)
\, , 
\end{align}
where $\lambda = \as \, \beta_0 \, \log N$, and $\widetilde {\mathcal{R}}_{\rm \scriptscriptstyle log} (\lambda) = e^{\frac{1}{a_s \beta_0} g_1(\lambda) + \dots}$ is the Sudakov exponential in $N$-space.
We then apply the Borel-prescription formula as for the case of a Mellin transform, see Eq.~(2.1.46) of \cite{Bonvini:2012sh}, thus obtaining
\begin{align}
\label{eq:Borel}
\frac{\rd \Rres^{\rm \scriptscriptstyle Borel}(\tau,W)}{\rd \tau}
\, = \,
\frac{1}{2 \pi i}
\oint
\frac{\rd \xi}{\xi} \, \,
\frac{\tau^{\,\xi-1}}{\Gamma(\xi)}
\int_0^W
\frac{\rd w}{\bar \alpha} \,
e^{-\frac{w}{\bar \alpha}} \,
{\widetilde {\mathcal R}}_{\rm \scriptscriptstyle log}
\Big(\!\!-\!\frac{w}{2 \xi} \Big)
\, ,
\end{align}
where $\bar \alpha = 2 \, a_s \, \beta_0$.
In Eq.~\eqref{eq:Borel} the integration path must enclose the origin $\xi = 0$, whereas $W \geq 2$ is a free parameter which can be used to estimate the ambiguity
%$\left(\frac{\Lambda_{\rm QCD}^2}{Q^2} \right)^{W/2}$ 
$\big(\Lambda_{\rm QCD}/Q\big)^W$ of the resummation procedure.
For our numerical implementation, we follow the procedure described in Ref.~\cite{Bonvini:2012sh} with $W=2$, and by integrating along the circle $\xi = r \, e^{i \theta}$, with $r = 1.3 + 2 \frac{w}{\bar \alpha} \ln (e^{1/\bar \alpha} - 1)$, to improve the numerical stability of the evaluation.

\begin{figure}
  \centering
  \includegraphics[width=0.5\textwidth]{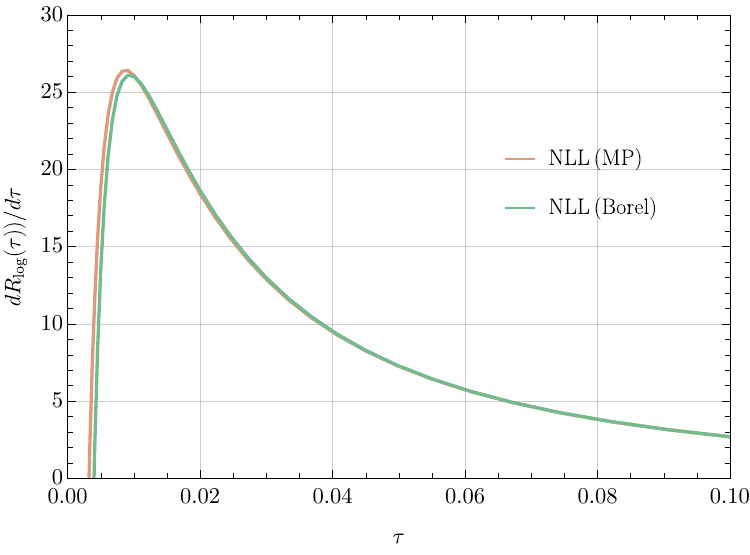}
  \caption{\label{fig:borel}
    Comparison between the NLL differential spectra for $\tau$ obtained by regularising the Landau pole with the minimal prescription (orange) and the Borel prescription (green).}
\end{figure}

In Fig.~\ref{fig:borel} we show a comparison at NLL between the results obtained using the minimal prescription and the Borel prescription.
The two results are in excellent agreement across most of the $\tau$ range, with differences visible only to the left of  the peak.

\bibliographystyle{JHEP}
\bibliography{thrustStudy.bib}
 
\end{document}